\documentclass[capt,preprint]{aastex}

\usepackage{amsmath,amssymb,graphicx,longtable,rotating,lscape}

\usepackage{natbib}
\let\new=\newcommand
\new{\be}{\begin{equation}}
\new{\ee}{\end{equation}}
\new{\disp}{\displaystyle}
\begin{document}

\shorttitle{X-ray variability in AGN}
%\title[Quasi-periodic behaviour in AGN light curves]{Analysis and modeling of quasi-periodic behaviour in light curves from Active Galactic Nuclei}
\title{X-ray variability and the inner region in AGN}

\author{P. Mohan\altaffilmark{1,\dagger} \& A.\ Mangalam\altaffilmark{1,\ddagger}}

\altaffiltext{1}{Indian Institute of Astrophysics, Sarjapur Road, Koramangala, Bangalore, 560034, India}

\email{prashanth@iiap.res.in\altaffilmark{\dagger}, mangalam@iiap.res.in\altaffilmark{\ddagger}}

\begin{abstract}
We present theoretical models of X-ray variability attributable to orbital signatures from an accretion disk including emission region size, quasi-periodic oscillations (QPOs) and its quality factor $Q$, and the emergence of a break frequency in the power spectral density shape. We find a fractional variability amplitude of $F_{var}\propto M^{-0.4}_{\bullet}$. We conduct a time series analysis on X-ray light curves ($0.3-10$ keV) of a sample of active galactic nuclei (AGNs). A statistically significant bend frequency is inferred in 9 of 58 light curves (16\%) from 3 AGNs for which the break timescale is consistent with the reported BH spin but not with the reported BH mass. Upper limits of $2.85 \times  10^7 M_\odot$ in NGC 4051, $8.02 \times  10^7 M_\odot$ in MRK 766 and $4.68 \times 10^7 M_\odot$ in MCG-6-30-15 are inferred for maximally spinning BHs. For REJ 1034+396, where a QPO at 3733 s was reported, we obtain an emission region size of $(6 - 6.5) M$ and a BH spin $a\lesssim$ 0.08. The relativistic inner region of a thin disk, dominated by radiation pressure and electron scattering is likely to host the orbital features as the simulated $Q$ ranges from $6.3 \times 10^{-2}$ to $4.25 \times 10^6$, containing the observed $Q$. The derived value of $Q \sim$ 32 for REJ 1034+396 therefore suggests that the AGN hosts a thin disk.
\end{abstract}


\section{Introduction}

X-ray emission and variability in radio-quiet Seyfert type 1s (Sy1s) and the narrow-line Seyfert 1s (NLS1s) are believed to be dominated by physical processes on the accretion disk in the context of unification models for active galactic nuclei (AGNs) \citep{1993ARA&A..31..473A,1995PASP..107..803U} as the observer line of sight intersects the disk. Variability on short timescales (few 1000 s to $<$ 1 day) in the X-rays can be studied with the availability of almost contiguous light curves for prominently visible AGNs with small sampling intervals from multiple space-based instruments such as {\em XMM Newton, Chandra, Suzaku} and others in the energy range of $0.1-100$ keV. 

Observed UV to X-ray spectra from these AGNs indicate the presence of a colder population of gas ($T$ $\leq$ 10$^5$ K) constituting the accretion disk sandwiched by another population at a much higher temperature, interpreted in terms of a layer of optically thin, thermally agitated relativistic electrons e.g. \cite{1991ApJ...380L..51H,1994ApJ...432L..95H}. If virial equilibrium is established in the corona populated by electrons with mass $m_e$, the temperature $kT = m_e c^2/r$ where $r = R/M$ is the radial distance, $R$, scaled in terms of the gravitational mass $M = G M_{\bullet}/c^2$ for a black hole (BH) of mass $M_{\bullet}$. For a non-rotating BH the innermost stable circular orbit is $r = r_{ISCO} = 6$ and $k \ T \sim$ 85.4 keV, the effective accretion energy per electron. Emission and variability in soft to hard X-rays are then expected to arise when disk-based seed photons (optical and UV photons) are up-scattered to higher energies through the inverse-Compton (IC) process e.g., \cite{2003PhR...377..389R}. For variability over a timescale $\Delta t$, the size of the emitting region can be at most $c \Delta t$. With a typical $\Delta t$ $\sim$ 1000 s to a few hours and for a black hole mass of $5 \times 10^6 M_{\odot}$, this corresponds to an emitting region from $\sim$ 40 $M$ to a few hundred $M$. The inferred location of the emission region from spectral modeling of the broad X-ray emission lines is typically even closer, between a few to a few tens of $M$  e.g., \cite{2011MNRAS.416.2725P}. Phenomenological models of variability in the optical/UV and X-ray emission from AGNs based on orbital signatures from the inner region of the disk around the central black hole (e.g. \citealt{1991A&A...246...21Z,1993ApJ...406..420M}) result in a power spectral density (PSD) with a power law whose slope is theoretically constrained between -1.4 and -2.1; this is well supported by observational studies in these wavelengths by (e.g. \citealt{2010ApJ...718..279G}). Hence we adopt this physical model in this paper.

In \cite{1999ApJ...524..667T}, an X-ray ($0.5-10$ keV) variability study was conducted with archival Advanced Satellite for Cosmology and Astrophysics (ASCA) data for 36 Seyfert 1 galaxies. The variance $\sigma_{\mathrm{rms}}$ was found to be related to the H$\beta$ FWHM as $\sigma_{\mathrm{rms}} \propto (\mathrm{FWHM} \ H\beta)^{-2.8}$, consistent with rapid variability and narrow lines leading to a small black hole mass. In the Rossi-X-ray-Timing-Explorer-based ($2-12$ keV), long-term X-ray variability studies spanning 3 years \citep{2001ApJ...547..684M} and 7 years \citep{2004ApJ...617..939M}, an anti-correlation between source luminosity and variability amplitude is measured on short timescales and is found to be in agreement with previous studies e.g. \cite{1986Natur.320..421B,1993ApJ...414L..85L} and it is suggested that this could be due to either a positive correlation between the luminosity and a break timescale or an inverse correlation between luminosity and overall amplitude. In \cite{2007MNRAS.375.1479S}, a one day study of simultaneous X-ray and optical variability was conducted for 8 nearby Seyfert 1 galaxies using XMM Newton. The X-ray variability amplitude was observed to be greater than that of the optical. In the optical, a maximum rms amplitude of 2.9\% is measured for NGC 3783. The rms amplitude for the X-ray light curves ranges between 2.0\% for Ark 120 and 47.6\% for NGC 4051. A cross correlation analysis between the optical and X-ray light curves does not indicate any significant correlation for three of the four objects showing a detectable optical variability, implying that re-processing of optical radiation may not be a dominant mechanism of production of the X-rays. In \cite{2011MNRAS.417.2426P}, a comparative study of the spectrum of 14 hard X-ray selected ($>$ 20 keV) NLS1 galaxies from the fourth International Gamma-Ray Astrophysics Laboratory's (INTEGRAL) Imager on Board the Integral Satellite (IBIS) catalog is conducted in the 0.3 keV to 100 keV energy range using data from the {\em XMM-Newton} satellite, the X-ray telescope aboard {\em Swift} satellite and INTEGRAL/IBIS. 
The study shows that NLS1 galaxies generally host low mass black holes with the calculated mass distribution of the objects peaking at $\sim$ 10$^7$ M$_{\odot}$. However, it must be noted that NLS1 black hole mass estimates consistently fall below the $M_{\bullet} - \sigma$ relation as indicated by previous studies e.g. \cite{2004ApJ...606L..41G}. Hence, the mass obtained from this method may be systematically underestimated by a small factor.

In \cite{2007ARA&A..45..441M}, a compilation of Seyfert galaxies indicating evidence for relativistically broadened X-ray emission lines is made. The study includes an analysis of X-ray data ($0.2-700$ keV) from {\em Chandra X-ray Observatory, XMM-Newton}, and {\em Suzaku}. The black hole spin for many of these objects, constrained through the spectral fit of the FeK$\alpha$ line is reported. Evidence supporting the spectral fit method of spin determination include the observation that the disk extends to the ISCO for a given spin $a$ above a threshold in the mass accretion rate, and arguments which imply that most radiation is from close to the ISCO with free falling material inside of it being fully ionized and thus emitting only weakly e.g. \cite{1998MNRAS.300L..11Y,2006ApJ...652.1028B}. In \cite{2010A&A...524A..50D}, an {\em XMM-Newton-}based ($2-10$ keV) statistical study of 149 Seyfert type 1 galaxies is conducted to collect evidence for a relativistically broadened Fe K$\alpha$ line. The main interpretations are drawn from a flux-limited sub-sample of 31 Seyfert type 1 galaxies. Strong evidence for a relativistically broadened Fe K$\alpha$ line is inferred for 36 \% of the flux-limited sources (11 of 31) and interpreted as a lower limit to the fraction of all possible sources which could provide such evidence. Inferences include an average line equivalent width of $\sim$ 100 eV, an average disc inclination of 28$^{\circ} \pm$ 5$^{\circ}$ and black hole spins, $a$, of 0.86$^{+0.01}_{-0.02}$ for MCG-6-30-15 and 0.74$^{+0.03}_{-0.04}$ for MRK 509. \cite{2012MNRAS.426.2522P} conducted a survey of the Fe K emission lines from archival {\em Suzaku} and {\em Swift} Burst Alert Telescope ($0.6-100$ keV) spectra of nearby ($z \leq$ 2) Seyfert 1 galaxies. After accounting for probable sources of emission, such as obscuring warm absorption clouds or a part of an outflow, the residual of the broad component is studied with a spectral fit. For the sample of 46 objects studied, 23 objects (50 \%) are found to require relativistic effects to account for the observed emission with a statistical significance $>$99.5\%. An average disk inclination towards the observer line of sight of 33$^{\circ}$ $\pm$ 2$^{\circ}$ is inferred from 20 objects and a maximally spinning BH ($a$ = 0.998) is ruled out for all objects with a confidence of 90\% with $a$ ranging between 0 and 0.80.

We present theoretical models of observational signatures in X-ray light curves due to orbital features which are likely to be present in the inner accretion disk in Section \ref{theory}. These include constraints on the emission region extent, the dynamic timescale, quasi-periodic oscillation (QPOs) attributed to an orbital period of these features, and properties which can be extracted from a detected QPO, followed by a discussion of the break frequency, its possible origin, and constraints on the emission region which can be inferred from the analysis of X-ray light curves. The data selection and reduction procedure is described in Section \ref{datasel}, followed by our analysis and significance testing procedure in Section \ref{analysis}. The results of the analysis are presented and discussed in Section \ref{results}. A brief summary of the important results are then presented in Section \ref{conclusions}.

\section{Theoretical models}
\label{theory}

In the following section, the radial coordinate $r$ is in units of the gravitational radius $M = G M_{\bullet}/c^2$ and the dimensionless BH spin $a$ is given by $J/(G M^2_{\bullet}/c)$, where $J$ is the angular momentum of the BH. We use geometrized units where $G$ = $c$ = 1.

\subsection{Emission region}
\label{emregion}

\begin{figure}
\centerline{\includegraphics[scale=0.25]{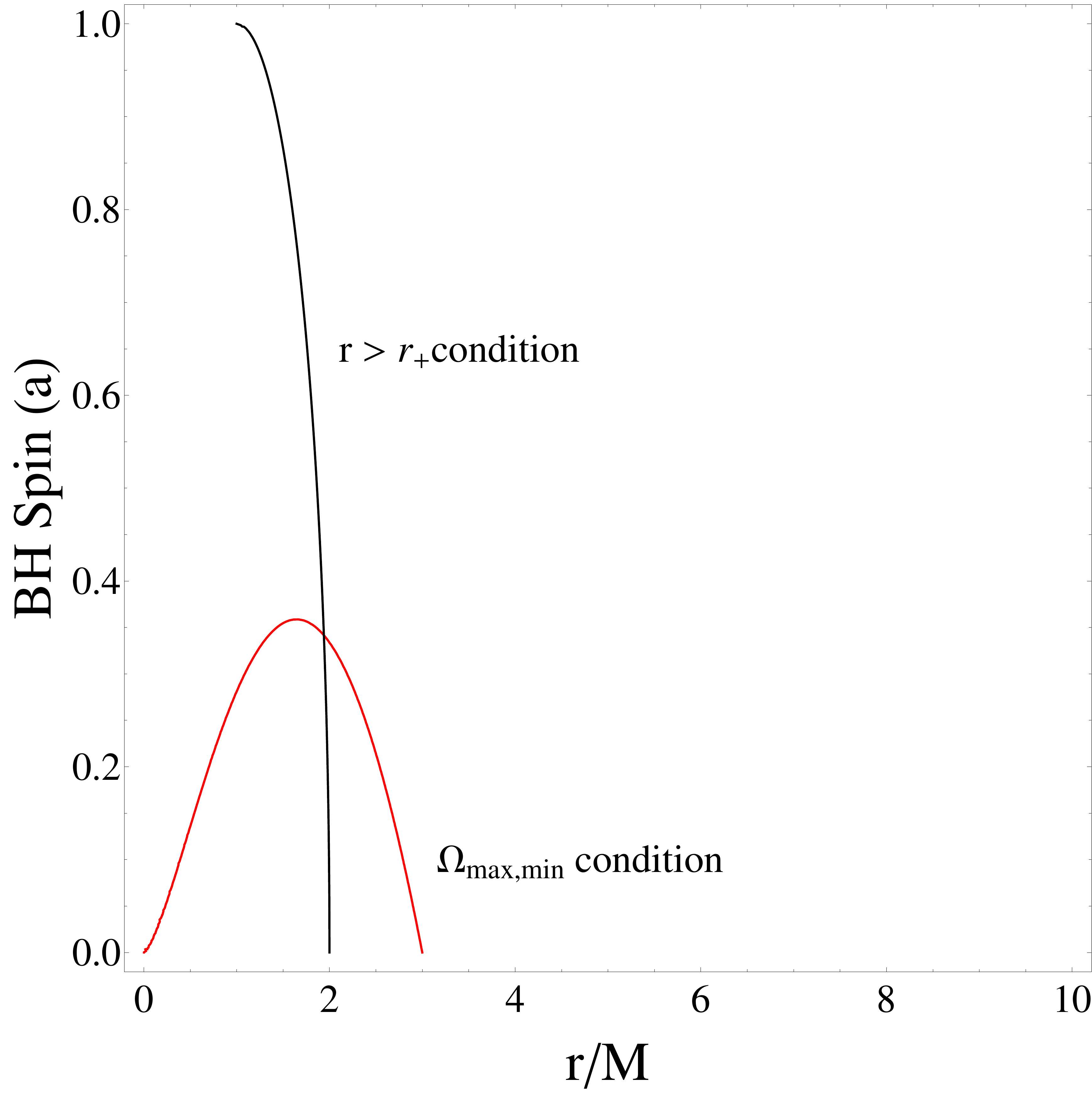}}
\caption[Constraints on minimum extent of emitting region]{Equality in the inequalities from Equations (\ref{eqn1}) and (\ref{eqn2}) are plotted as functions of $r$ and $a$. The region to the right of both the contours is the allowed region for emission.}
\label{racont}
\end{figure}

Emission could at most arise from sources near the innermost stable circular orbit (ISCO) for a Keplerian disk around the SMBH. We can place constraints on the minimum size of the emitting region, the spin of the SMBH, and its mass using certain conditions. The constant spin $\Omega$ in a rotating metric must obey $\Omega_{\rm min} < \Omega <\Omega_{\rm max}$ where 
\begin{equation}
\Omega_{\rm max, min} ={-g_{t \phi} \pm (g^2_{t \phi} -g_{tt} g_{\phi \phi})^{\frac{1}{2}} \over g_{\phi \phi}}.
\label{omega}
\end{equation}
For the Kerr case this leads to the condition that (see the Appendix \ref{appA}),
\begin{eqnarray}
4 a^4 r+8 a^4+2 a^3 r^{\frac{7}{2}}+8 a^3 r^{\frac{5}{2}} +8 a^3 r^{\frac{3}{2}}+a^2 r^5 \nonumber \\
-a^2 r^4 -2 a^2 r^3 +2 a r^{\frac{11}{2}}+4 a r^{\frac{9}{2}}+r^7-3 r^6 > 0,
\label{eqn1}
\end{eqnarray}
where $r$ is the radial coordinate and $a$ is the dimensionless BH spin. In addition, the emitting region must lie outside the event horizon $r_+$ of the hole, given by,
\begin{equation}
r > r_+=1+\sqrt{(1-a^2)}.
\label{eqn2}
\end{equation}
%The constraints above on the emission region and spin of the black hole along with information on the red-shift $z$ of the AGN can be used in Equation (\ref{time}) to place upper limits on the SMBH mass $M_{\bullet}$. 
A plot representing the above conditions is presented in Figure \ref{racont}.

\subsection{Dynamical timescale}
\label{theorydyn}

Dynamic or orbital processes yield the shortest characteristic timescale over which inhomogeneities, such as flares, in disks and jets cause changes, and the timescale is given by $t_\phi \sim r/v_\phi$. The Keplerian angular frequency of a test particle in circular motion around a Kerr black hole, eg. \cite{1972ApJ...178..347B}, is given by,
\begin{equation}
\Omega = \frac{2 \pi}{T} = \frac{1}{r^{3/2}+a}.
\label{kepler}
\end{equation}
$T$ is the orbital periodicity associated with the angular frequency and $r$ is the radial distance from the black hole with spin $a$. Using the above expression in C.G.S. units, $T$ can be written as
\begin{eqnarray}
\label{time}
T &=& 2 \pi (r^{3/2}+a) \ (1+z) \ G M_{\bullet}/c^3 \ s \\ \nonumber
 &=& 30.93 (r^{3/2}+a) \ m_6 \ (1+z) \ s,
\end{eqnarray}
where $T$ has been corrected to include the cosmological red-shift factor $z$ and the black hole mass is scaled in terms of the solar mass with $m_6 = M_{\bullet}/(10^6 M_\odot)$. If there is a statistically significant short timescale QPO in X-ray light curves, then this could be caused by orbiting inhomogeneities constituting the bulk flowing plasma on the accretion disk. This material undergoes a gradual radial drift and moves towards the black hole, leading to quasi-periodicity.

A measure of the evolution of the periodicity $T$ or the orbital frequency is the quality factor of a QPO, $Q = \Omega/\Delta \Omega$ where $\Delta \Omega$ is the change in the angular frequency caused by the radial drift of the orbiting material during each orbit. %The quantity, $\disp{\frac{\Delta T}{T} = \frac{\Delta \Omega}{\Omega} = Q^{-1}}$. 
The quality factor can be expressed in terms of the physics in the inner accretion disk \citep{2010A&A...521A..15A},
\begin{equation}
Q^{-1} = \frac{\Delta T}{T} = \frac{\Delta \Omega}{\Omega} = \frac{1}{\Omega} \frac{d \Omega}{dr}  \frac{dr}{dt}.
\end{equation}
As we consider $\Delta T$ to be the change in the periodicity due to radial drift over one orbit, $\Delta T = d \phi/\Omega = 2 \pi/\Omega$. Then,
\begin{equation}
Q^{-1} = \frac{2 \pi}{\Omega^2} \frac{d\Omega}{dr}  \frac{u^r}{u^t}
\end{equation}
where $u^r$ and $u^t$ are the four-velocity components of the orbiting flow. For a bulk flow of material in Kerr geometry, $\disp{u^r = \gamma_r \beta_r \sqrt{\frac{\Delta}{r^2}}}$ and $\disp{u^t = \sqrt{\frac{A}{\Delta r^2}} \gamma_{\phi} \gamma_{r}}$, obtained for a bulk flow in radial motion with a Lorentz factor $\gamma_r$ and a velocity $\beta_r$ as well as orbital motion with a Lorentz factor $\gamma_\phi$ on the disk as viewed by an observer in the local non-rotating frame \citep{1998ApJ...498..313G}, $A = (r^2+a^2)^2-a^2 \Delta \sin^2 \theta$ and $\Delta = r^2+a^2-2 M r$ are quantities used in the expression for the Kerr metric. Thus,
\begin{equation}
\frac{u^r}{u^t} = \frac{\beta_r \Delta}{\sqrt{A} \gamma_\phi};
\end{equation}
using this $Q$ can then be written as
\begin{equation}
Q^{-1} = \frac{2 \pi}{\Omega^2} \frac{d\Omega}{dr}  \frac{\beta_r \Delta}{\sqrt{A} \gamma_\phi}.
\end{equation}
$Q$ thus depends on the radial distance $r$, the polar angle defining the plane of motion of the emitting source $\theta$ and the black hole spin $a$. The azimuthal Lorentz factor is $\gamma_\phi = \sqrt{1-(v^{\phi})^2}$ where $v^{\phi}$ is the azimuthal orbital velocity of the bulk flow.  This is given by the expression $\disp{v^{\phi} = \frac{A \sin \theta}{\Sigma} \frac{\Omega-\omega}{\sqrt{\Delta}}}$, where $\Sigma = r^2+a^2 \cos^2 \theta$ and $\disp{\omega = 2 a r/A}$ is the rotational angular frequency due to the frame dragging effect of the spinning black hole. If we assume that the flow is along the equatorial plane ($\theta = \pi/2$) and for Keplerian angular velocity, $\Omega$  as in Equation\ (\ref{kepler}), then the quantity $Q$ depends on the radial distance $r$, the spin of the black hole, and the radial velocity of the bulk flow $\beta_r$,
\begin{equation}
Q = \frac{1}{3 \pi r^{1/2}} \frac{\sqrt{A}}{\beta_r \Delta} \left(1-\frac{(A-2 a r (r^{3/2}+a))^2}{\Sigma^2 \Delta}\right)^{-1/2}.
\label{Qfac}
\end{equation}

%A simple estimate of $\beta_r$ can be made. The local sound speed in the medium is $c_s \sim \sqrt{k T/m_e}$ for a gas composed of electrons of mass $m_e$ obeying the ideal gas equation. For a typical disk temperature $T \sim 10^5$ K as in the inner disk, $c_s \sim 1.23 \times 10^6$ m/s. As $v_r \ll c_s$, $\beta_r = v_r/c \ll$ 0.004. 
Using the \cite{1973blho.conf..343N} model for a general relativistic thin disk with a viscosity parameter $\alpha$, we use a mass accretion rate scaled to the Eddington accretion rate $\dot{m} = \dot{M}/\dot{M}_{\mathrm{Edd}}$. The scaled mass used here, $m_6$ makes the quantity relevant in the regime of supermassive black holes hosted by AGNs. The radial velocity, $\beta_r$, for the disk can be written for three important regimes: the plunge region dominated by gas pressure and electron-scattering-based opacity between $r_{ISCO}$ and the black hole horizon, the edge region dominated by gas pressure and electron scattering based opacity at or close to $r_{ISCO}$ and an inner region dominated by radiation pressure and electron-scattering-based opacity at small radii comparable to $r_{ISCO}$. In the following equations, $\beta_r = \beta_r (r,a,m_6,\dot{m})$ is given by \citep{2012MNRAS.420..684P}
\begin{equation}
\beta_{r,Plunge} = -\left((\mathcal{C}^{-1}_o \mathcal{G}^2_o \nu-1) + 1.9 \times 10^{-3} \alpha^{1/4} m^{-1/4}_6 \dot{m}^{1/2} r^{-7/8}_o \mathcal{C}^{-5/4}_o \mathcal{D}^{-1}_o \mathcal{G}^2_o \nu \right)^{1/2},
\end{equation}

\begin{equation}
\beta_{r,Edge} = - 7.1 \times 10^{-6} \alpha^{4/5} m^{-1/5}_6 \dot{m}^{2/5} r^{-2/5} \mathcal{B}^{4/5} \mathcal{C}^{-1/2} \mathcal{D}^{3/10} \Phi^{-3/5},
\end{equation}

\begin{equation}
\beta_{r,Inner} = - 124.416 \ \alpha \ \dot{m}^2 \ r^{-5/2}  \mathcal{A}^{2} \mathcal{B}^{-3} \mathcal{C}^{-1/2} \mathcal{D}^{-1/2} \mathcal{S}^{-1} \Phi.
\end{equation}

The parameters $\mathcal{A}$, $\mathcal{B}$, $\mathcal{C}$, $\mathcal{D}$, $\mathcal{G}$, $\mathcal{S}$, $\nu$ are functions of $r$ and $a$, and $\Phi$ is a function of $r$, $a$, $m_6$ and $\dot{m}$. These parameters express the thin disk structure equations in Kerr geometry and are given in \cite{2012MNRAS.420..684P}. 

A study of long-term timescales associated with optical B-band variability \citep{2004MNRAS.347...67S} offers an interpretation in terms of the thermal timescale for re-adjustment of plasma flow on the accretion disk. The viscosity parameter $\alpha$ is constrained in the narrow range $0.01 - 0.03$ for $L/L_{\mathrm{Edd}} = 0.01 - 1$ in the study. A compilation of several observational studies \citep{2007MNRAS.376.1740K} related to dwarf nova outbursts, outbursts by soft X-ray transients, AGN variability on thermal timescales, proto-stellar disks and FU Orionis outbursts gives $\alpha$ in the range $0.1 - 0.4$. The scaled mass accretion rate $\dot{m}$ ranges between 0.01 and 0.3 for a thin disk beyond which the disk becomes ``puffed'' up and transitions to the slim disk regime eg. \cite{2013LRR....16....1A}. 

\begin{figure}
\centerline{\includegraphics[scale=0.3]{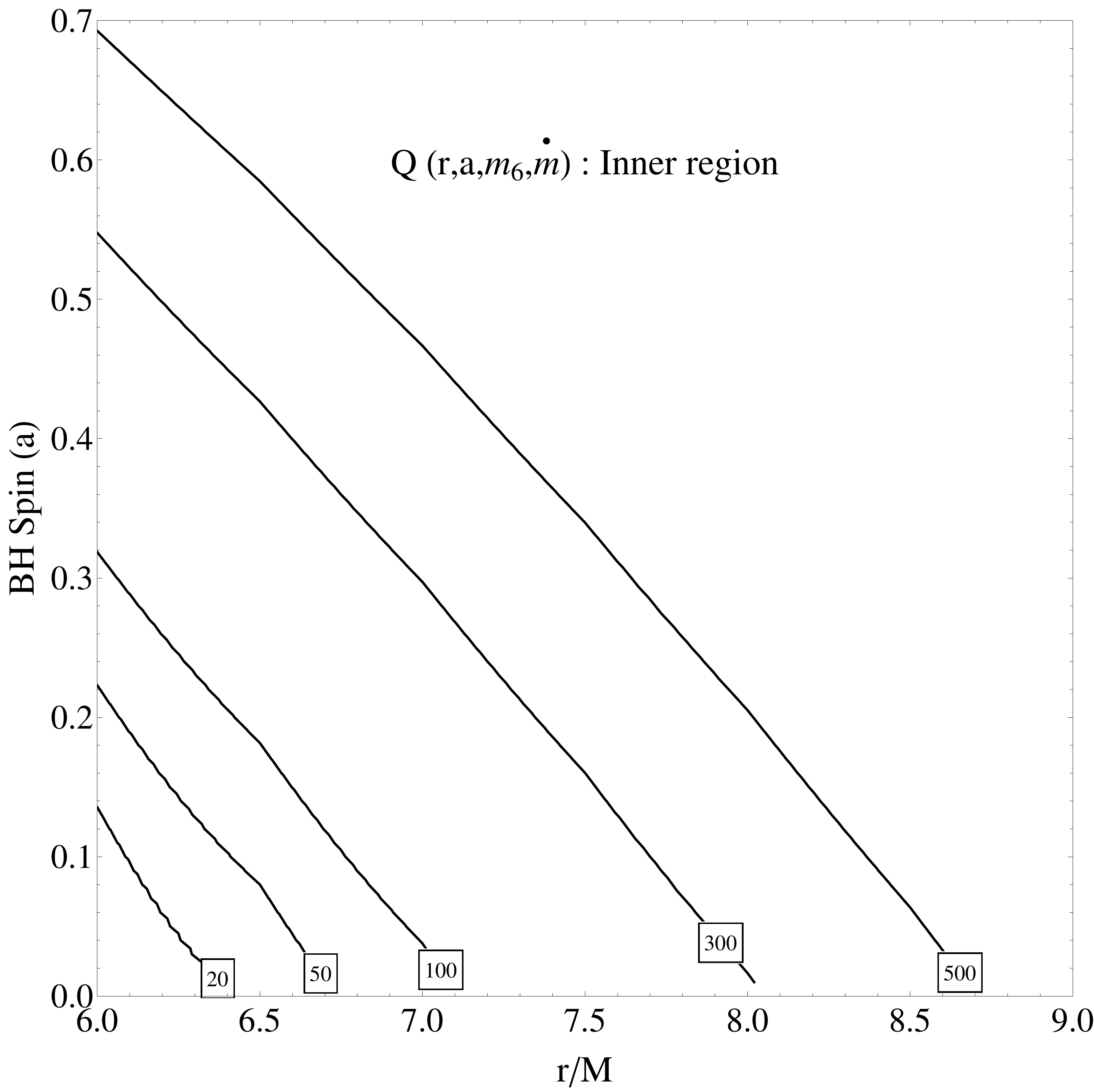}}
\caption[Quality factor $Q$]{Contours of $Q(r,a,m_6,\dot{m})$ for $\alpha = 0.1$, $r = 6 M-9 M$, $m_6 = 1 - 9$ and $\dot{m} = 0.01-0.3$. Shown here are the $Q$ contours in the physically relevant (study of AGN QPOs) range of $20 - 500$. The contours of $Q$ are plotted only for those $\beta_r$ $<$ 1. These values of $\beta_r$, and hence $Q$, satisfy are consistent with the inequality conditions expressed in Equations (\ref{eqn1}) and (\ref{eqn2}).}
\label{QfacInner}
\end{figure}

Based on the above estimates of the range of $\alpha$ and $\dot{m}$, we conducted a set of three simulations to determine the range of $r$, $a$ and $Q$ in the plunging, edge and inner regions for $\alpha$ = 0.01, 0.1 and 0.4. The scaled BH mass $m_6$ was incremented in steps of 1 in the range 1 - 10 and the mass accretion rate $\dot{m}$ in steps of 0.02 in the range $0.01 - 0.3$ for each of these regions. We also ensured that the choices of $r$ and $a$ obeyed the inequalities expressed in Equations (\ref{eqn1}) and (\ref{eqn2}). The results of these simulations are presented in Table \ref{Qsims}. For the plunging region, we infer $Q = 0.15 - 0.23$, realizable for $\beta_r = 0.46 - 0.99$ with $r = 3.37 M - 5.90 M$ and a BH spin $a = 0.03 - 0.5$. At these small radii, instabilities which formed at larger radii are expected to be smoothed out as they drift in. Coupled with highly relativistic radial plunge velocities, it is then expected that their orbital signatures are very weak at best, thus giving rise to a range of low $Q$. For the edge region, we infer $Q = 2761 \ (\alpha = 0.4) - 4.25 \times 10^6 \ (\alpha = 0.01)$, realizable for $\beta_r = 1.96 \times 10^{-8} \ (\alpha = 0.01) - 8.39 \times 10^{-5} \ (\alpha = 0.4)$ with $r = 2.80 - 5.95$. The contours of $Q=Q(r,a,m_6,\dot{m})$ are plotted in Figure \ref{QfacInner} for the inner region for which we infer $Q = 6.3 \times 10^{-2} \ (\alpha = 0.4) - 3.52 \times 10^4 (\alpha = 0.01)$, which is realizable for $\beta_r = 1.36 \times 10^{-6} - 0.99$ with $r\geq r_{ISCO} - 20 M$.

\begin{table}
\centerline{{\bf $Q$ simulation results}}
\scriptsize{
\begin{tabular}{|l|l|l|l|l|l|l|}
\hline
Viscosity & ($r$,$a$,$\beta_r$) & $Q$ & ($r$,$a$,$\beta_r$) & $Q$ &($r$,$a$,$\beta_r$) & $Q$ \\
$\alpha$  & Plunge            & Plunge& Edge            & Edge& Inner            & Inner \\ \hline
0.01     & 3.37 - 5.90       & 0.15 - 0.23 & 2.80 - 5.95 & 4.38 $\times$ 10$^{3}$ & 5.00 - 9.00 & 8.5 $\times$ 10$^{-2}$ \\ 
         & 0.03 - 0.50       &             & 0.82 - 0.99 &  - 4.25 $\times$ 10$^{6}$ & 0.24 - 0.99 & - 3.52 $\times$ 10$^{4}$   \\
         & 0.46 - 0.99       &             & 1.96 $\times$ 10$^{-8}$ - 5.30 $\times$ 10$^{-5}$ &         & 1.36 $\times$ 10$^{-5}$ - 0.99 &\\ \hline
0.1     & 3.37 - 5.90       & 0.15 - 0.23 & 2.80 - 5.95 & 3.28 $\times$ 10$^{3}$ & 5.00 - 9.00 & 6.9 $\times$ 10$^{-2}$ \\ 
         & 0.03 - 0.50       &             & 0.82 - 0.99 & - 6.96 $\times$ 10$^{5}$   & 0.32 - 0.99 & - 3.52 $\times$ 10$^{3}$ \\
         & 0.46 - 0.99       &             & 1.24 $\times$ 10$^{-7}$ - 7.05 $\times$ 10$^{-5}$ &         & 1.36 $\times$ 10$^{-5}$ - 0.99 &\\ \hline
0.4 & 3.37 - 5.90       & 0.15 - 0.23 & 2.80 - 5.95 & 2.76 $\times$ 10$^{3}$ & 5.00 - 9.00 & 6.3 $\times$ 10$^{-2}$ \\ 
         & 0.03 - 0.50       &             & 0.82 - 0.99 & - 5.85 $\times$ 10$^{6}$  & 0.45 - 0.99 &  - 8.80 $\times$ 10$^{3}$ \\
         & 0.46 - 0.99       &             & 3.75 $\times$ 10$^{-7}$ - 8.39 $\times$ 10$^{-5}$ &         & 5.45 $\times$ 10$^{-5}$ - 0.99 &\\
\hline
\end{tabular}}
\caption{Results of simulations to illustrate the clear demarcation in the $Q$ values for the plunge, edge and inner regions of a relativistic thin disk. The radial velocity $\beta_r$ is in units of the speed of light $c$, the radius $r$ is in units of $G M_{\bullet}/c^2$ and the dimensionless spin $a$ is in units of $J/(G M^2_{\bullet}/c)$.}
\label{Qsims}
\end{table}

The evaluated $Q$ values in the above simulations show a distinct demarcation between the physically relevant regions. In addition, this demarcation is robust against increasing $\alpha$. The range of $Q$ measured from observations is a few tens in AGNs, eg. $\sim$ 32 in REJ 1034+396 measured in \cite{2014JApA..MMC}. The inner region case is thus best suited to describe orbital processes as the simulated $Q$ lies in this observationally relevant range. This would also indicate that the QPO phenomenon is likely to arise only from orbital features in the inner region of the thin disk.

If a bulk flow is radially drifting inward towards the central BH and if its radial velocity $\beta_r$ is very low ($\sim$ 0.0001), then any orbital feature present on the disk is likely to exist for a large number of cycles. Also, the change in orbital frequency $\Delta \Omega$ is very small for this $\beta_r$, in turn giving rise to a high $Q$, and hence a sharp QPO feature. If $\beta_r$ is large (e.g. 0.1), then there is insufficient time for the QPO to develop as the radial bulk motion is very fast, leading to a broad QPO. In many cases, this would merge with the continuum of the PSD shape, rendering it difficult to statistically identify. This could be a possible explanation for the absence of a QPO detection in the light curves of radio quiet AGN. Thus, if a QPO arises due to the emission from orbital features on the disk, the emission radius $r$ and black hole spin $a$ can be constrained if the black hole mass is determined accurately by techniques such as reverberation mapping. Once the AGN luminosity is fixed through an analysis of its broadband spectrum, $\dot{m}$ is known. The free parameters which remain include $r$ and $a$. One can then solve Equations (\ref{time}) and (\ref{Qfac}) to obtain $r$ and $a$. 

\subsection{Break frequency and the region of emission}
\label{brkfreq}
\begin{figure}
\centerline{\includegraphics[scale=0.3]{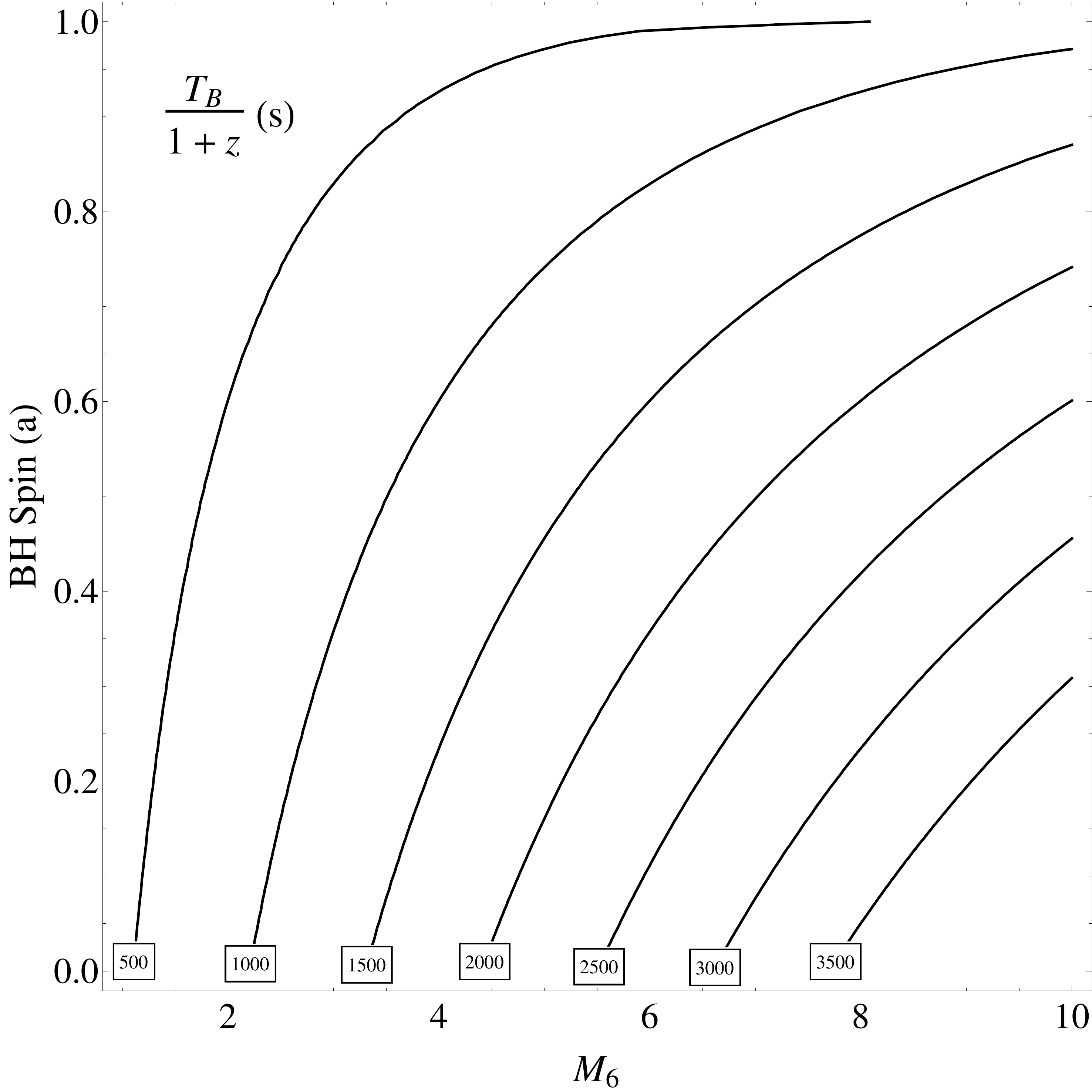}}
\caption[$T_B$ for $r$ = 6 $M$, $m_6$ = 1 - 10 and $a$ = 0 - 0.998]{Contours of the break timescale $T_{B} = T_{B}(r,a)$. The timescales indicated in the plot range between $\sim$ 500 s and 3500 s for the black hole mass in units of $m_6$ ranging between 1 and 10 and the spin $a$ ranging between 0 and 0.998. The timescales indicated in the plot are chosen such that they can be used in the direct comparison for $T_{B}$ inferred from short timescale X-ray light curves where it is expected to range between a few 100 s and a few 1000 s.}
\label{brkfig}
\end{figure}

A break frequency in analyzed light curves is characterized by a change in the power-law slope of the PSD in the frequencies lower than the break (flatter with slope ranging between 0 and $-1$) and in the frequencies higher than the break (steeper with slope ranging between $-1$ and $-2$). The PSD shape is thus expected to steepen at high frequencies ($\geq$ 10$^{-4}$ Hz). This could occur due to emitting bulk flow on the disk making a transition from tightly bound orbits on the disk to a free-falling inward spiral towards the central black hole upon crossing the inner edge of the accretion disk. For Keplerian angular velocity, $\disp{\Omega = \frac{M}{r^{3/2}+a}}$ (as in Equation (\ref{kepler})), the break timescale $T_B$ is related to the location of $r_{ISCO}$ and is given by
\begin{equation}
T_B = 30.93 \ m_6 (1+z) f(a) \ {\rm s},
\label{brktime}
\end{equation}
where 
\begin{equation}
f(a) = (r^{\frac{3}{2}}_{ISCO}(a)+a)
\end{equation}
and the quantity $r_{ISCO} (a)$ is given by \citep{1972ApJ...178..347B},
\begin{equation}
 r_{ISCO} = 3+Z_2(a) -[(3-Z_1(a)) (3+Z_1(a)+2 Z_2(a)]^{\frac{1}{2}},
\end{equation}
where
\begin{eqnarray} 
Z_1(a) = 1+(1-a^2)^{\frac{1}{3}} [(1+a)^{\frac{1}{3}}+(1-a)^{\frac{1}{3}}] \\ \nonumber
Z_2(a) = (3 a^2+Z^2_1(a))^{\frac{1}{2}}.
\end{eqnarray}
Thus, the break timescale $T_B$, corrected to include the cosmological red-shift factor, is dependent on the black hole mass scaled in terms of the solar mass, $m_6 = M_{\bullet}/(10^6 M_\odot)$ and the black hole spin $a$. The contours of $T_B(M_{\bullet},a)$ are plotted in Figure \ref{brkfig}. A strong consistency condition arises if a light curve indicates both a QPO and a break timescale. The QPO can be used to determine $r$, $a$ and $\dot{m}$ if the black hole mass is known through the use of Equations (\ref{time}) and (\ref{Qfac}). From this analysis, we must obtain $r$ and $a$ that satisfy the inequalities Equations (\ref{eqn1}) and (\ref{eqn2}) in order for the description to be physically relevant and describe signatures from orbital features on the disk.

\section{Data selection and reduction}
\label{datasel}

Our preliminary sample of study included a set of radio-quiet AGNs where the broad FeK$\alpha$ 6.4 keV fluorescence line is well detected \citep{2007ARA&A..45..441M}. As this line is expected to be emitted from the inner accretion disk e.g. \cite{2004MNRAS.348.1415V}, and the observer line of sight intersects the accretion disk in a canonical AGN model from unification schemes \citep{1993ARA&A..31..473A,1995PASP..107..803U}, these AGNs can be used to study disk-based variability. From this set, we chose a sample of AGNs with BH mass measured from either reverberation mapping or the $M- \sigma$ relation and spin estimates derived from Fe K$\alpha$ line shapes as reported in literature. 

From the {\em XMM Newton} online archives, a set of 58 EPIC-PN detector-based light curves for 16 Seyfert galaxies in the soft X-ray band ($0.3-10$ keV) are extracted and reduced for the analysis using SAS 12.0. Reduction, light curve extraction and pre-processing of the light curves are carried out as follows.

The light curve for the energy range $>$ 10 keV was first observed and manually truncated to a good time interval (GTI) by eliminating regions of the light curve dominated by proton flaring events. Then, we placed a circular aperture of size ranging from 35$^{''}$ to 45$^{''}$ around the object depending on the specific data set. The light curve of the object was then extracted with a time bin size of 100 s in the GTI for the energy range $0.3-10$ keV. By placing a circular aperture of the same size as the object slightly away (about 200$^{''}$) from the source to prevent contamination by source photons, we extracted the light curve of the background in the same energy range with the same bin size in the GTI. The size of the apertures around the object and the background are tailored for each data set in order to avoid dark regions or patches with no counts. The object light curve with time bin size of 100 s was then obtained by subtracting the background from the source. Where small data gaps ($<$ 10 points) were present in an object light curve, caused due to the removal of flaring portions, we performed a linear interpolation in this region and used these estimates to obtain the final analyzable light curve with time bin size of 100 s, similar to the procedure followed in \cite{2012A&A...544A..80G}.

For all the extracted light curves, the fractional excess rms variability amplitude $F_{\mathrm{var}}$ \citep{2003MNRAS.345.1271V} is determined from
\begin{equation}
F_{\mathrm{var}} = \sqrt{\frac{\sigma^2-\hat{\sigma}^2_{\mathrm{err}}}{\mu^2}},
\end{equation}
where $\sigma_{\mathrm{err},k}$ is the uncertainty or measurement error for each point of the light curve, $x(t_k)$, evaluated at times $t_k$, and $\disp{\hat{\sigma}^2_{\mathrm{err}} = \frac{1}{N} \sum^{N}_{k=1} \sigma_{\mathrm{err},k}}$ is the mean square error, with $\sigma^2$ the variance and $\mu$ the mean of the light curve evaluated from the points of the light curve. 

The AGNs studied here and their properties, including their black hole mass $M_{\bullet}$ and spin $a$ obtained from the literature and the average $F_{\mathrm{var}}$ calculated from their light curves, are presented in Table \ref{AGNsummary}.

\section{Data analysis}
\label{analysis}

\subsection{Periodogram and PSD models}

Our analysis involves a fit to the measured periodogram with parametric models to infer the actual PSD shape. The normalized periodogram is given by,
\begin{equation}
P(f_j)=\frac{2 \Delta t}{\mu^2 N} |F(f_j)|^2
\end{equation}
where $\Delta t$ is the time step size, $\mu$ is the mean of an evenly sampled, mean-subtracted light curve $x(t_k)$ of length $N$ points, and $|F(f_j)|$ is its discrete Fourier transform evaluated at frequencies $f_j = j/(N \Delta t)$ with  $j = 1, 2, .., (N/2-1)$. With this normalization, the integrated periodogram gives the fractional variance of the time series \citep{2003MNRAS.345.1271V} and $P(f)$ is in units of (rms/mean)$^2$ Hz$^{-1}$. The periodogram is then fit with two PSD models, the power-law and the bending power-law models, based in order for a comparison with previous works eg. \cite{2012A&A...544A..80G}. The power law model is given by
\begin{equation}
I(f_j) = A f^{\alpha}_j+C.
\end{equation}
The free parameters in this model include the amplitude $A$, the red-noise slope $\alpha$ often found in the range of $-1$ and $-2.5$ in the periodograms of AGNs, and the constant Poisson noise $C$. This model fits optical/ultra-violet and X-ray data reasonably well and could represent broadband variability across a wide range of Fourier frequencies due to collective and random processes on the disk.

The bending power law model is given by
\begin{eqnarray}
I(f_j) = A f^{-1}_j \left(1+(f_j/f_b)^{-\alpha-1}\right)^{-1}+C.
\end{eqnarray}
The free parameters in this model include the amplitude $A$, the red-noise slope $\alpha$, the bend frequency $f_{b}$, and the constant Poisson noise $C$. The model is used here to represent the shape of the expected PSD with a break discussed in Section \ref{brkfreq} and is a special case of the generalized bending knee model \citep{2004MNRAS.348..783M}.

\subsection{Fit procedure, confidence intervals and model selection}

The statistical procedure involved in the fit to the periodogram uses the maximum likelihood estimator (MLE) to determine model fit parameters. The likelihood and log-likelihood functions e.g. \cite{2013MNRAS.433..907E,2014PhDT...PM,2014JApA..MMC} are defined as
\begin{equation}
L(\theta_k) = \prod^{(n-1)}_{j=1} \frac{1}{I(f_j,\theta_k)} e^{-P(f_j)/I(f_j,\theta_k)},
\end{equation}
%S(\theta_k) &=& -2 \ln (L(\theta_k)) \\ \nonumber
\begin{equation}
 S(\theta_k) = - 2 \sum^{n-1}_{j=1}(\ln(I(f_j,\theta_k))+P(f_j)/I(f_j,\theta_k)),
\end{equation}
where $L(\theta_k)$ is the likelihood function, $I(f_j,\theta_k)$ are parametric models, $\theta_k$ are the parameters of $I(f_j,\theta_k)$ to be estimated and $P(f_j)$ is the periodogram of the light curve. Determining $\theta_k$, which minimize $S$, yields the maximum likelihood values. 

Confidence limits are determined in a manner similar to that described in \cite{2002nrc..book.....P} for the $\Delta \chi^2$ method. The log-likelihood $S$ is determined for a large number of combinations of the parameters $\theta_k$ for a given model. The global minimum $S_{\mathrm{min}}$ is determined. The parameter combination yielding this is unique and yields the best fit model. A space of $\Delta S$ is then constructed using $\Delta S = S_i - S_{\mathrm{min}}$ where the $S_i$ corresponds to each unique combination of the parameters $\theta_k$. $\Delta S$ are approximately $\chi^2_k$ distributed with the degrees of freedom $k$ corresponding to the number of parameters in the model. Confidence intervals are thus set based on the cumulative distribution function of the $\chi^2_k$ distribution. These would correspond to a given value of $\Delta S$. For example, $\Delta S  = 2.71$ for 90\% confidence intervals for one parameter. The $\Delta S$ for a confidence interval $p$ in a space of $k$ dimensions is in general in terms of the inverse of the regularized gamma function $Q^{-1}$ and is given by $2 Q^{-1} (\nu/2,0,p)$ \citep{2013MNRAS.433..907E} where $\nu$ corresponds to the degrees of freedom ($\nu = 2$ when the evaluation is carried out at all frequencies other than the Nyquist frequency). Thus, $\Delta S$ determined from the parameter combinations within a set confidence intervals are grouped. The parameter ranges that they span can thus be determined using which one can specify the confidence intervals of all parameters $\theta_k$ used in a given model. This procedure is repeated for all parametric models. In each case, the best-fit parameter values and the confidence interval of each of those parameters is determined.

Model selection is carried out using the Akaike information criteria (AIC) and the likelihood (L), e.g. \cite{2004..Likelihood}. The AIC is a measure of the information lost when a model is fit to the data. The model with least information loss (least AIC) is taken as the best fit.  As the likelihood is proportional to the probability of the success of the model in effectively describing the PSD shape, we use this in the model selection. The AIC and likelihood are defined by
\begin{align}
AIC&=S(\theta_k)+2 p_k, \\ \nonumber
\Delta_i&=AIC_{\mathrm{min (model \ i)}}-AIC_{\mathrm{min(null)}}, \\ \nonumber
L(\mathrm{model \ i|data})&=e^{-\Delta_i/2},  \\ \nonumber
\end{align}

where $p_k$ is a penalty term and is the number of parameters $\theta_k$ used in the model, and $L(\mathrm{model \ i|data})$ is the likelihood of model $i$ given the data. $AIC$ and $L$ are determined for each competing model. Models with $\Delta_i\leq 2$ can be considered close to the best fit, those with $4 \leq \Delta_i \leq 7$ are considerably less supported, and those with $\Delta_i > 10$ and $RL > 150$ cannot be supported \citep{2004..Likelihood}.

Significance levels are used to identify outliers in the periodogram which could either be strictly periodic components or quasi-periodic oscillations (QPOs). As the quantity $P(f_j)/I(f_j)$ is expected to follow the $\chi^2_2$ distribution, the area under the tail of the probability density function of the $\chi^2_2$ distribution (which is equivalent to a gamma density $\Gamma(1,1/2)= \exp{(-x/2)}/2$) gives the probability $\epsilon$ that the power deviates from the mean at a given frequency and is measured in units of standard deviation given by $\gamma_{\epsilon}$. For $K$ number of trial frequencies used to construct the best-fit model, we can specify a $(1-\epsilon) \ 100 \%$ confidence limit on $\gamma(f_j)$ \citep{2005A&A...431..391V} by means of
\begin{equation}
\gamma_\epsilon=-2 \ln[1-(1-\epsilon)^{\frac{1}{K}}] .
\end{equation}
After specifying an $\epsilon$ such as 95 \% or 99.99 \%, $\gamma_\epsilon$ is calculated, and the quantity $\gamma_\epsilon$ is multiplied with $I(f_j)$ to give the significance level. The statistical identification of a QPO as a true feature in a light curve can be supplemented through the use of a suite of essential time series analysis techniques in a consistent manner with a search strategy \citep{2014JApA..MM}. The analysis suite includes the periodogram, the Lomb Scargle periodogram, the multi-harmonic analysis of variance periodogram and the wavelet analysis. The search strategy helps in the identification as well as confirmation of the QPO and, in addition, provides properties of the QPO such as the number of cycles and when it was present during the observation duration.

\section{Results and discussion}
\label{results}

\begin{figure}
\centerline{\includegraphics[scale=0.25]{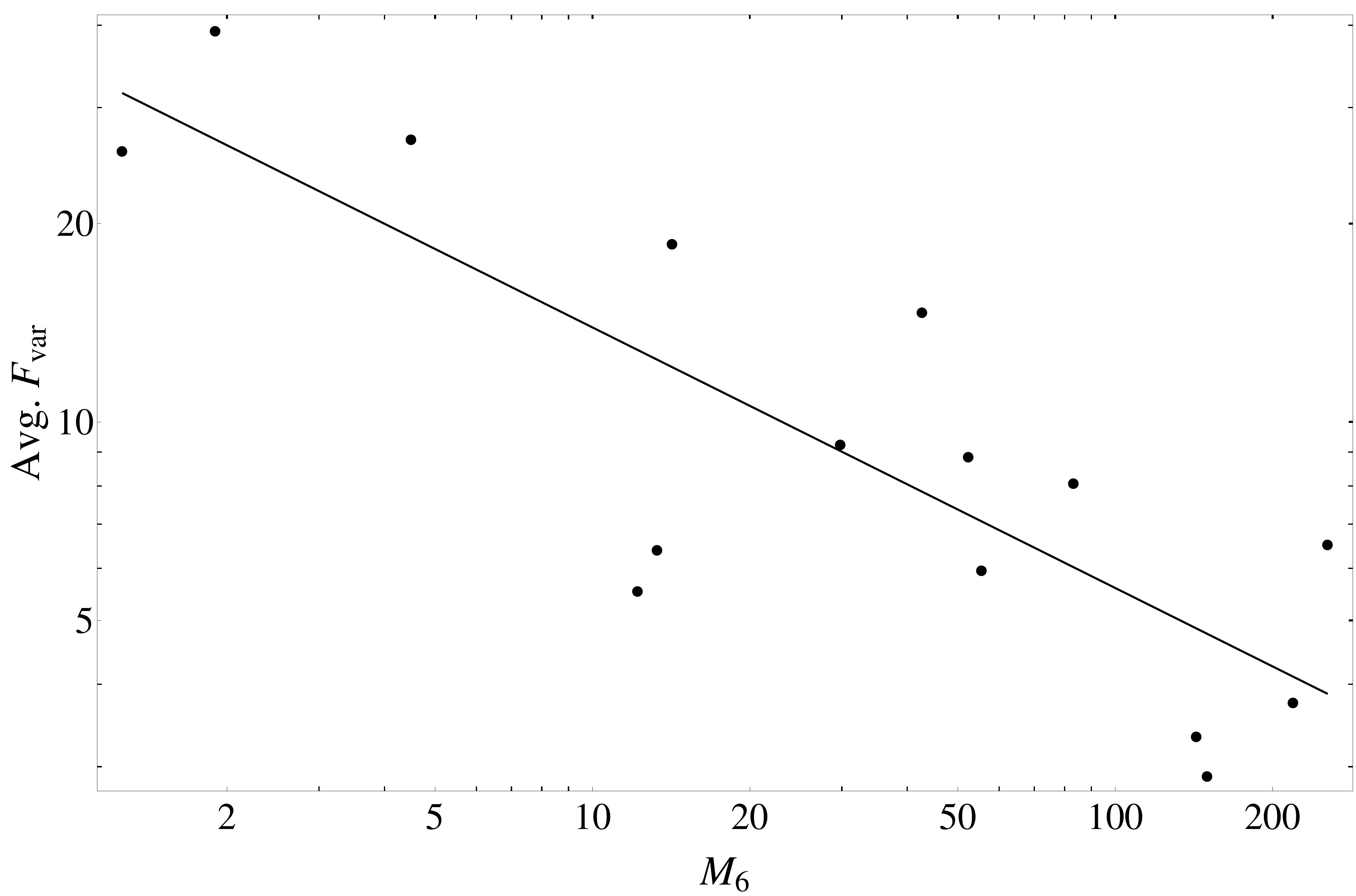}}
\caption[$F_{\mathrm{var}}$ vs $M_{\bullet}$ for including all objects]{Linear fit in the log$-$log space to the $F_{\mathrm{var}}$ vs. $M_{\bullet}$ data for Sy1 and NLSy1 galaxies.}
\label{fvarplot2}
\end{figure}

We perform a linear fit to the $F_{\mathrm{var}}$ versus $M_{\bullet}$ data in the log$-$log space with and without the results of the NLSy1 population (MRK 335, NGC 4051 and MRK 766) due to suggestions from literature e.g. \cite{2004MNRAS.350L..26N} which expect $F_{\mathrm{var}} \propto M_{\bullet}^{-0.5}$ only when Sy1s are considered. The average $F_{\mathrm{var}}$ ranges between 2.91 (ARK 120) and 26.89 (MCG-6-30-15) for a reduced sample of 13 objects. We removed another point for Q 0056-363 as its black hole mass estimate is uncertain and the quoted value has not been confirmed using the multiple mass measurement procedures. The fit to the average $F_{\mathrm{var}}$ versus $M_{\bullet}$ data in the log$-$log space for 12 objects yields a slope of $-0.33 \pm 0.12$, which is flatter compared to the expected $-0.5$. With the inclusion of the data from the NLSy1s, the fit for 15 objects yields a steeper slope of $-0.40 \pm 0.08$ plotted in Figure \ref{fvarplot2}, within the error bars of previous studies.

%MC simulations based significance testing of detected peaks in the LSP, MHAoV and wavelet analysis periodograms do not indicate any statistically significant QPO in any of the analyzed data sets. 

\begin{figure}
\centerline{\includegraphics[scale=0.3]{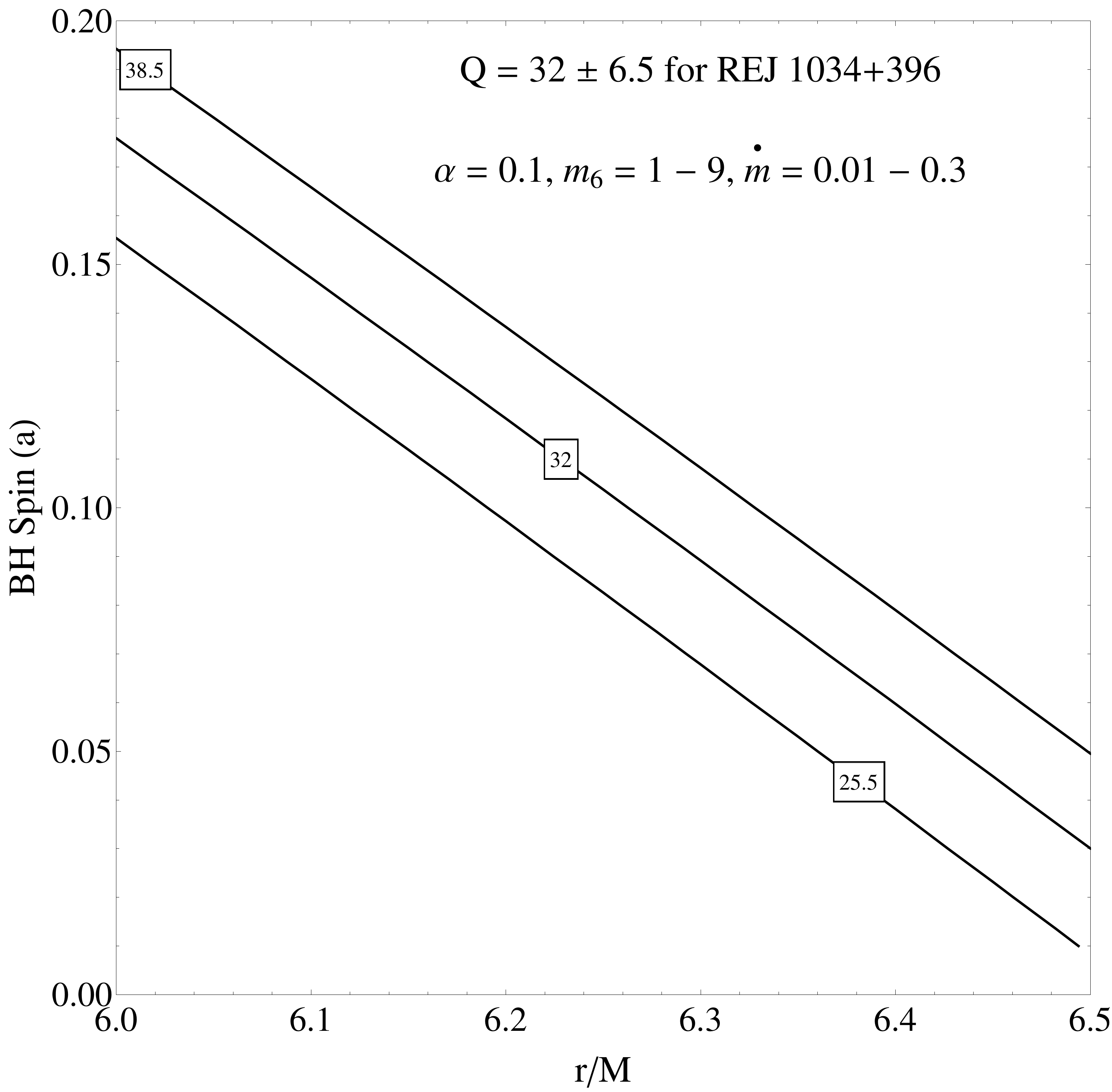}}
\caption[Quality factor $Q$]{Simulations of $Q(r,a)$ were performed for $\alpha = 0.1$, $r = 6 M - 16 M$, $m_6 = 2 - 7$ and $\dot{m} = 0.01 - 0.3$ to obtain the observed $Q$ = 32 $\pm$ 6.5 contours. We obtain $r = 6 M - 6.5 M$ and $a$ $\leq$ 0.08 for $\dot{m}$ $\leq$ 0.3.}
\label{QfacInnerREJ}
\end{figure}

We can illustrate the usefulness of the developed theoretical formalism in Section \ref{theorydyn} by applying it to the case of the NLSy1 galaxy REJ 1034+396 ($z = 0.034$) where a QPO peaked at (3733 $\pm$ 140) s was reported by \cite{2008Natur.455..369G}. A study of the X-ray reverberation in REJ 1034+396 indicates that the flux from the soft X-rays ($<$ 1 keV) and that from the hard X-rays ($>$ 3 keV) lags behind the flux in the intermediate X-ray band ($1-3$ keV) by $\sim$ 290 s at higher spectral frequencies ($<$ 3.5 $\times$ 10$^{-4}$ Hz) interpreted as the relativistically smeared reflected emission dominating the soft and hard X-ray flux \citep{2011MNRAS.418.2642Z}. The lags are inferred to originate from the inner region at a few gravitational radii with the QPO in this data set inferred to originate from the corona. The emission in the $0.3-10$ keV X-ray band is thus likely to be dominated by orbital processes in the inner accretion disk. In \cite{2014JApA..MMC}, we performed the PSD analysis of this light curve. The power law with a Lorentzian QPO was determined to be the best-fit PSD shape (QPO significance $>$ 99.94 \%) with an amplitude $R$ of 0.05 $\pm$ 0.01 and a quality factor $Q$ of 32.0 $\pm$ 6.5. Broadband spectrum modelling yields a black hole mass of $(2 - 10) \times 10^6 M_{\odot}$ and an accretion rate, $\dot{m}$, of $0.3 - 0.7$ \citep{2001ApJ...550..644P}. The black hole mass is estimated to be (4$^{+3}_{-2}$) $\times$ 10$^6$ M$_\odot$ from the relation between the excess variability amplitude and the black hole mass \citep{2010ApJ...710...16Z}. Using the latter mass estimate of $m_6$ = 4$^{+3}_{-2}$ along with $T = (3733 \pm 130)$ s, we estimate $r$ to lie in the range 6.25 $M$ and 15.70 $M$ for $a$ ranging between 0 and 0.998 using Equation \ref{time} outside $r_{ISCO}$. Then, we constrain $r$ and $a$ further using $Q = 32.0 \pm 6.5$ and $\alpha = 0.1$. Assuming that the AGN hosts a thin accretion disk, we use Equation (\ref{Qfac}) with the radial inflow velocity for a Novikov-Thorne \citep{1973blho.conf..343N} disk as modified and applied to the inner disk region \citep{2012MNRAS.420..684P}. Using $\alpha = 0.1$, we simulate $Q = Q(r,a)$ with $r = 6 M - 16 M$, $a = 0.01 - 0.998$, $m_6 = 2 - 7$ and $\dot{m} = 0.01 - 0.3$, and we obtain a radial inflow velocity $\beta_r = 3.8 \times 10^{-5} - 0.99$. By constraining allowable values of $Q = 32 \pm 6.5$, we were able to reduce constraints on $r = 6 M - 6.5 M$ and $a \leq 0.08$, plotted in Figure \ref{QfacInnerREJ}. The inflowing material could also attain relativistic radial velocities and accretion could become advectively dominated or it could lead to the formation of a slim disk which can lead to higher luminosities, even super Eddington, in which case we would require the corresponding prescription for the radial inflow velocity. In the present situation, $Q$ lies in the range of simulated $Q$ for the inner region close to $r_{ISCO}$. This strongly suggests that REJ 1034+396 hosts a relativistic thin disk.

\begin{figure}
\centerline{\includegraphics[scale=0.22]{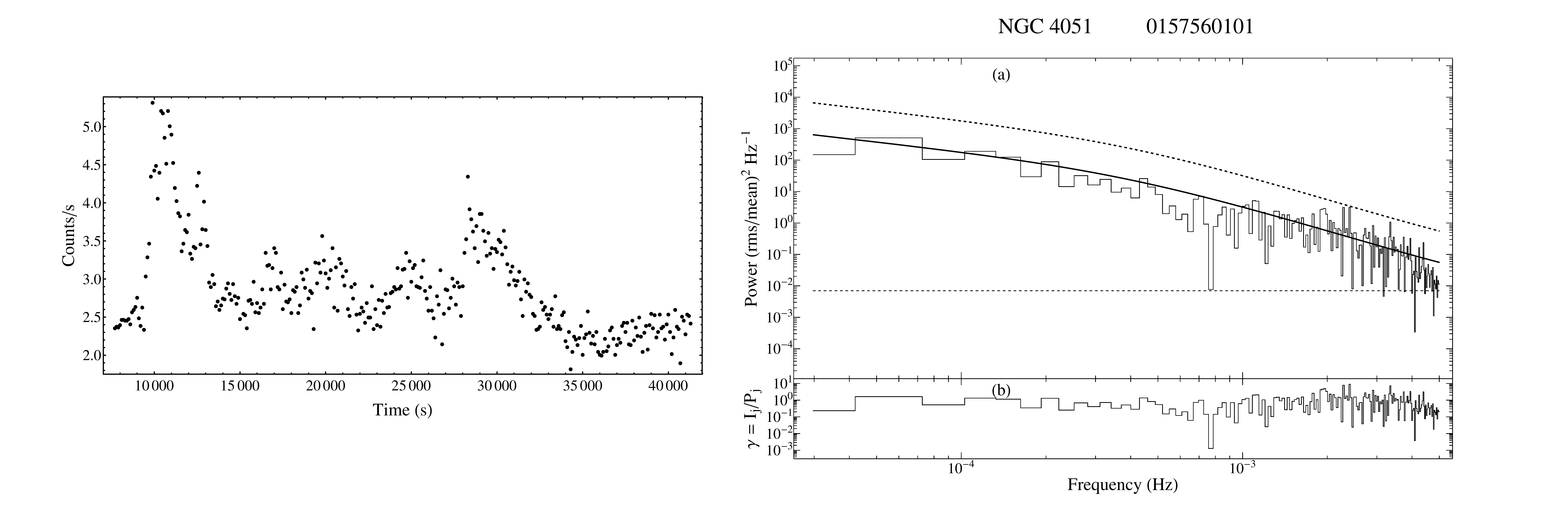}}
\centerline{\includegraphics[scale=0.22]{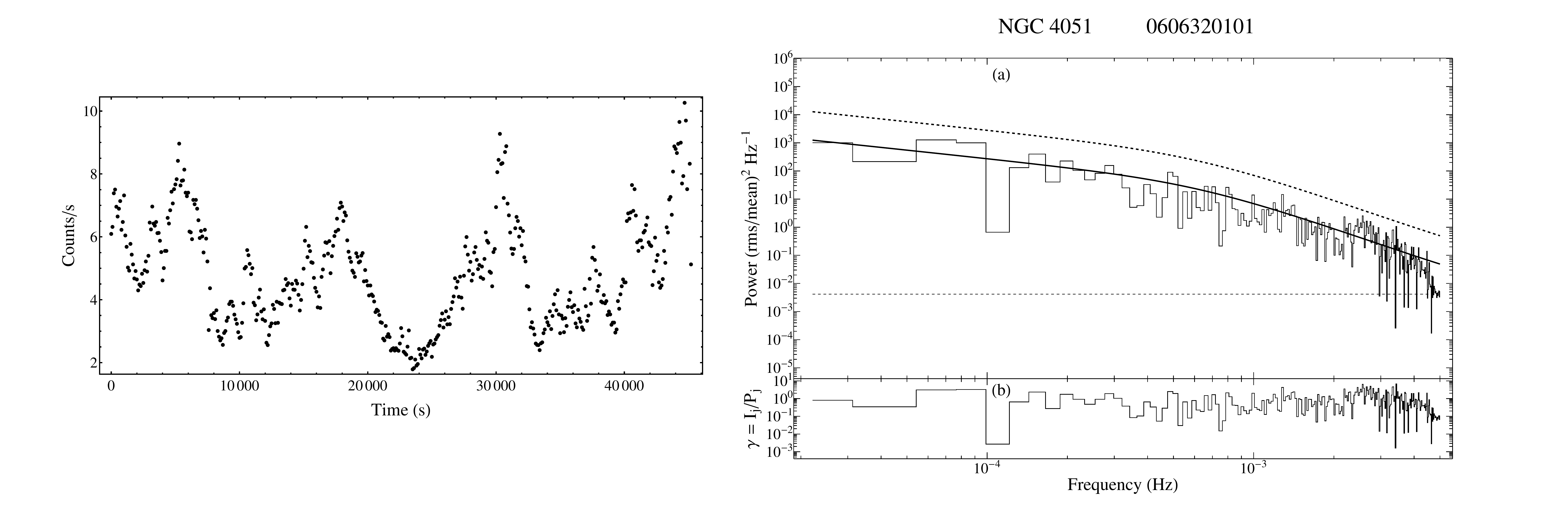}}
\centerline{\includegraphics[scale=0.22]{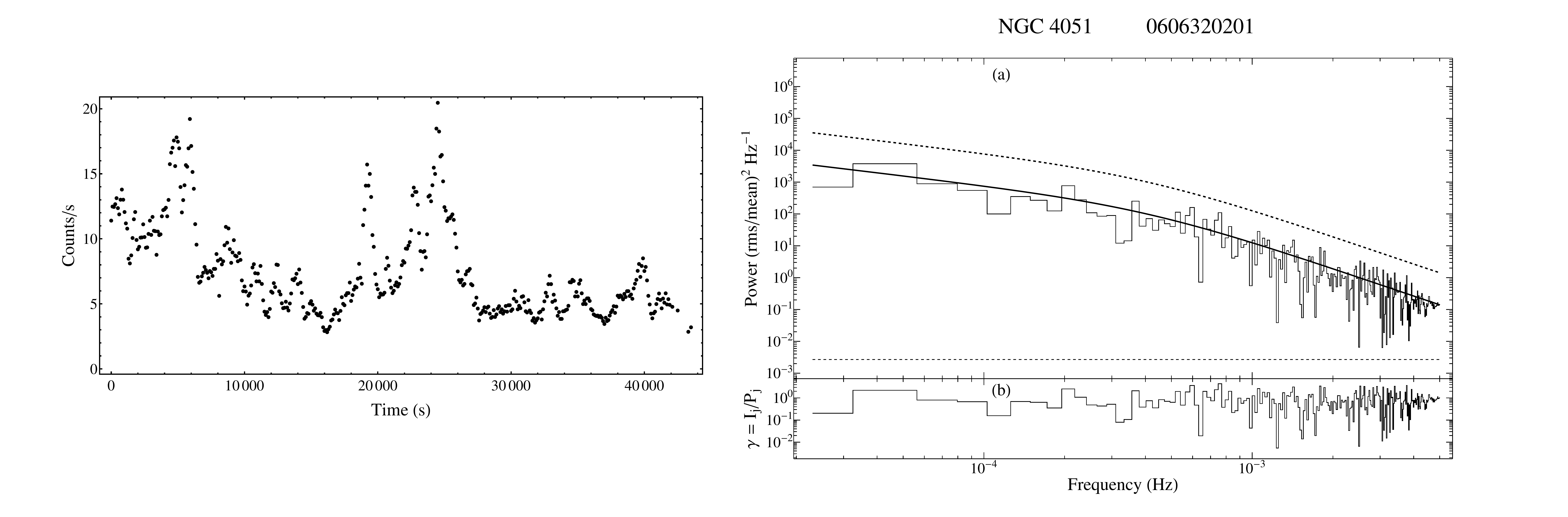}}
\caption{Left plots: light curve of AGNs indicating a PSD with a bending power law. Right plots: panel (a) in all the above plots is the periodogram of the light curves showing a statistically significant bend frequency. The fit portion includes a 99 \% dashed confidence level plotted above it. The horizontal dashed line is the white noise level. Panel (b) is the ratio $\gamma = I_j/P_j$ which is the fit residual.}
\label{Xrayplots1}
\end{figure}

\begin{figure}
\centerline{\includegraphics[scale=0.22]{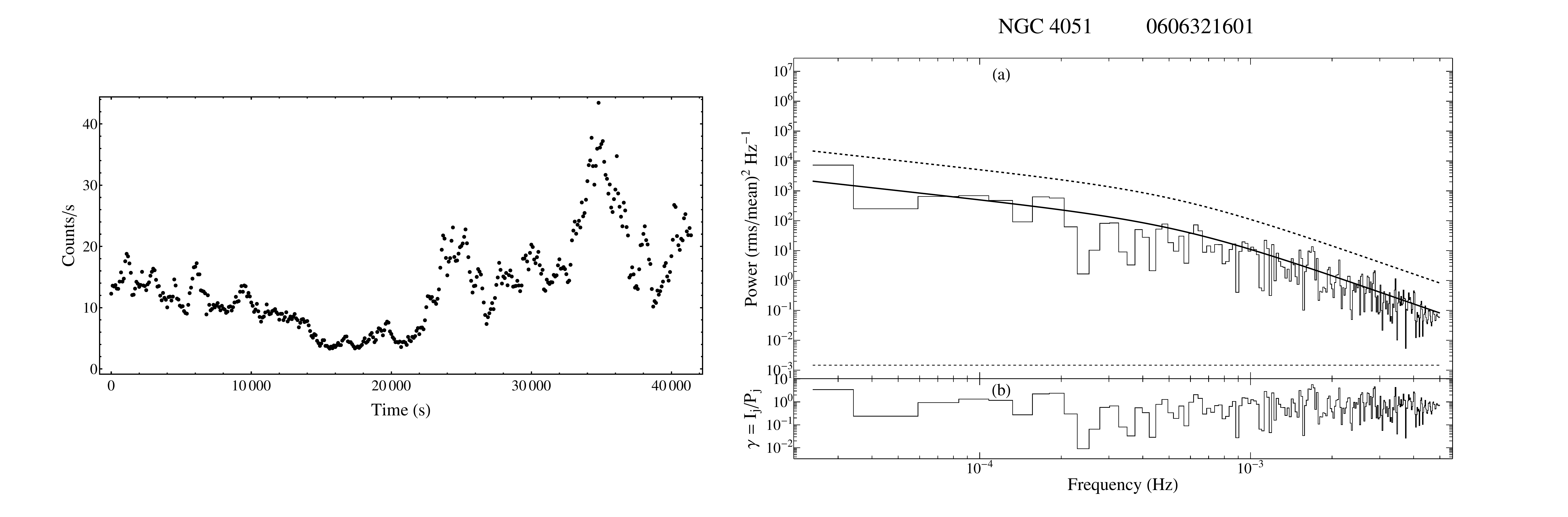}}
\centerline{\includegraphics[scale=0.22]{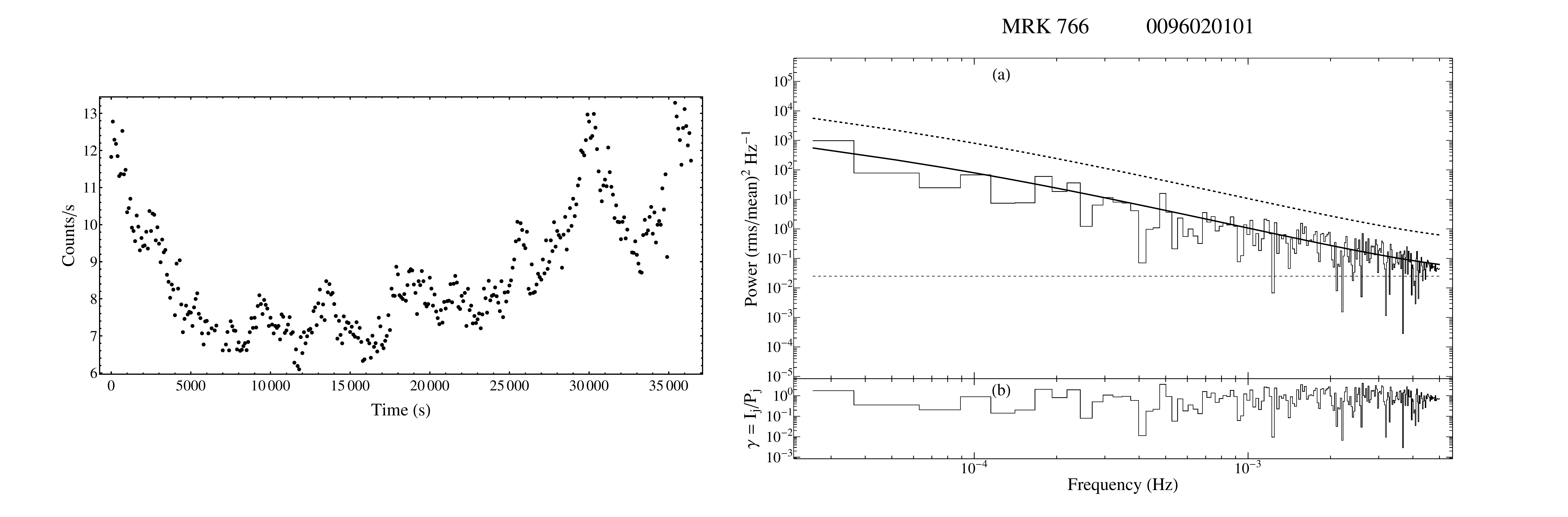}}
\centerline{\includegraphics[scale=0.22]{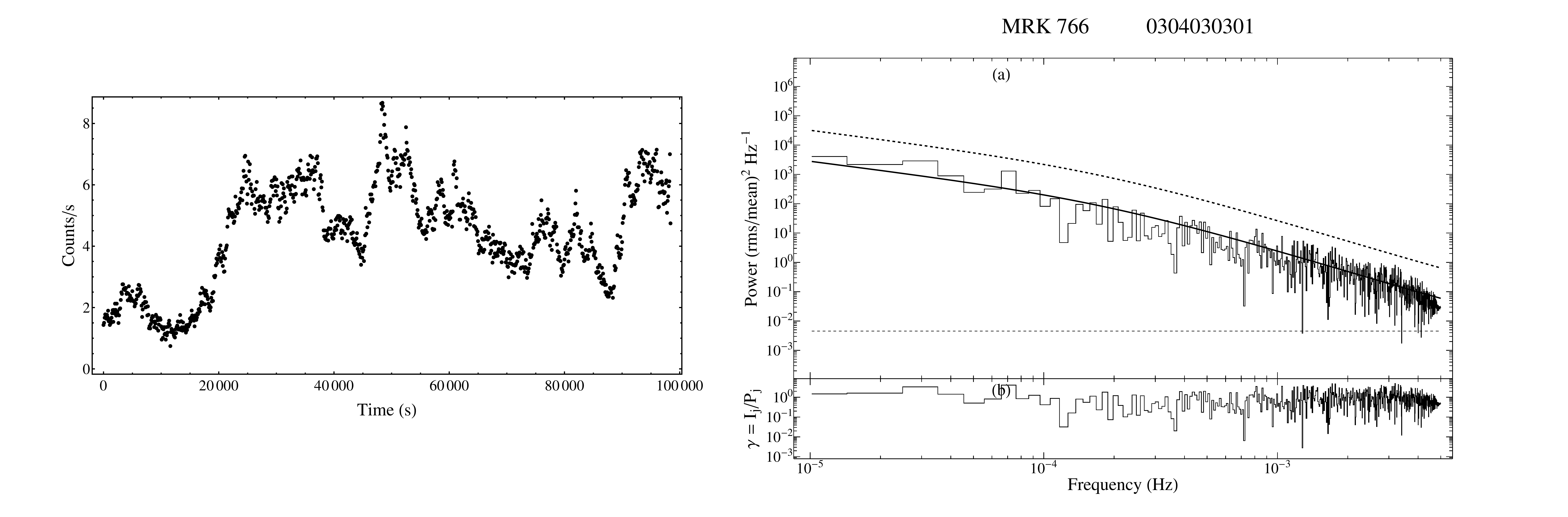}}
\caption{Left plots: light curve of AGNs indicating a PSD with a bending power law. Right plots: panel (a) in all the above plots is the periodogram of the light curves showing a statistically significant bend frequency. The fit portion includes a 99 \% dashed confidence level plotted above it. The horizontal dashed line is the white noise level. Panel (b) is the ratio $\gamma = I_j/P_j$ which is the fit residual.}
\label{Xrayplots2}
\end{figure}

\begin{figure}
\centerline{\includegraphics[scale=0.22]{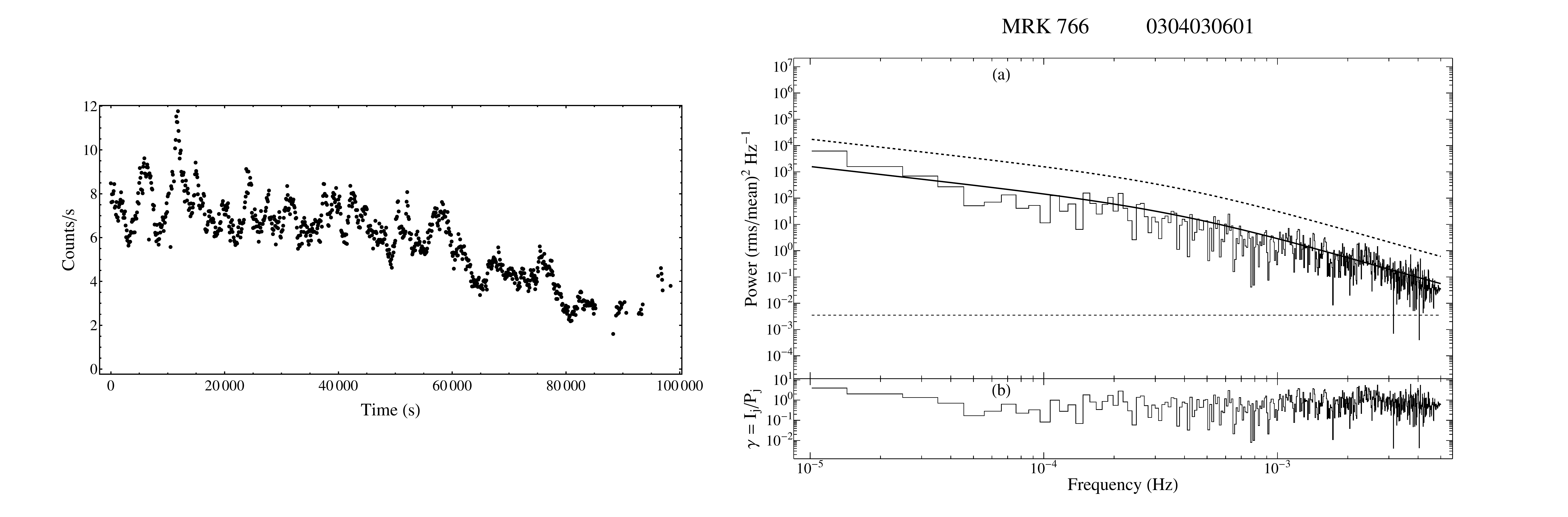}}
\centerline{\includegraphics[scale=0.22]{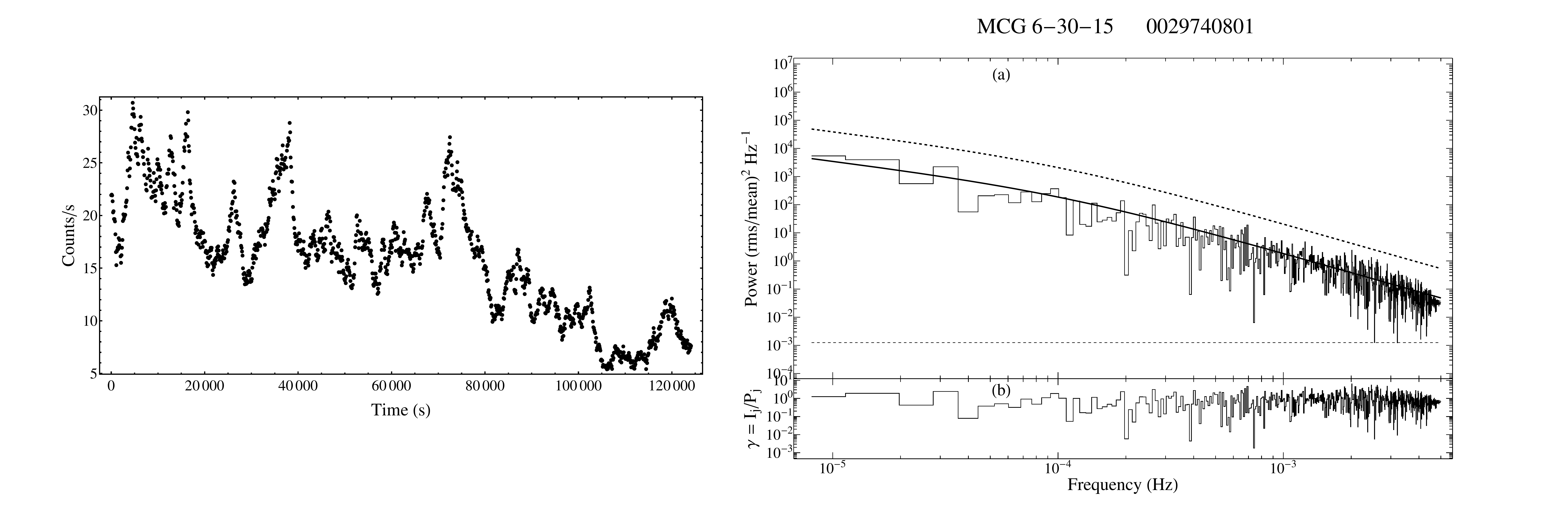}}
\centerline{\includegraphics[scale=0.22]{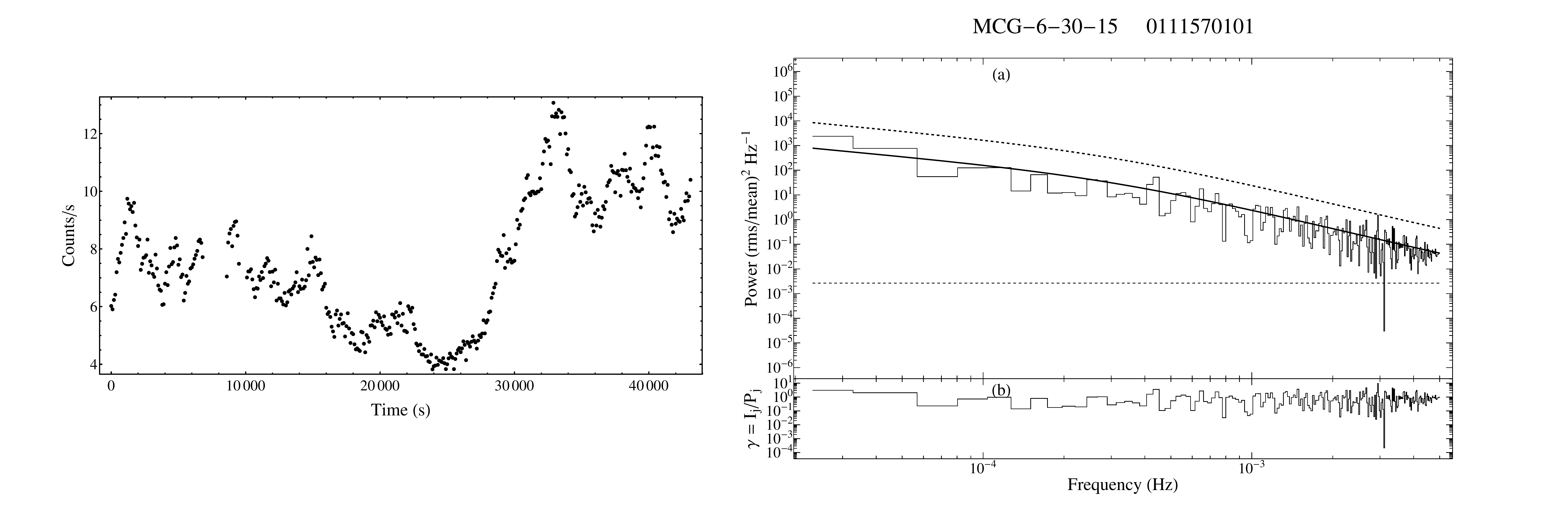}}
\caption{Left plots: light curve of AGNs indicating a PSD with a bending power law. Right plots: panel (a) in all the above plots is the periodogram of the light curves showing a statistically significant bend frequency. The fit portion includes a 99 \% dashed confidence level plotted above it. The horizontal dashed line is the white noise level. Panel (b) is the ratio $\gamma = I_j/P_j$ which is the fit residual.}
\label{Xrayplots3}
\end{figure}

Table \ref{pgrambestfit} summarizes the results from the periodogram analysis of each light curve. $F_{var}$ for these light curves ranges between 2.30\% and 53.48\% indicating a moderate to strong variability. The power-law model of the PSD is a good fit in 36 of the 58 (62\%) light curves. The bending power-law model of the PSD is a good fit in 9 of the 58 (16\%) light curves. In the remaining 13 light curves, the model selection procedure was unable to provide a distinct difference to choose one model over the other, or the values obtained for the bend frequency were lower than $-3.00$ which is likely to arise from noise fluctuations, or higher than $-4.00$ which could arise from long term trends. In the nine light curves indicating a bending power law, the bend timescale ranges between between $1820^{+269}_{-235}$ s (NGC 4051) and $6026^{+4207}_{-2478}$ s (MCG 6-30-15). In case there were multiple bend timescales indicated for an AGN, we took the weighted mean of all these instances to represent the bend frequency for that AGN. We then apply the theoretical break frequency model from Section \ref{brkfreq} to these light curves. The results of the time series analysis of the above light curves where a $T_B$ is inferred are plotted in Figures \ref{Xrayplots1}$-$\ref{Xrayplots3}. 

We make use of the BH mass $M_{\bullet}$ and spin $a$ ranges as quoted in the literature and presented in Table \ref{AGNsummary} for the analysis. Since $T_{B} = T_{B} (M_{\bullet},a)$ from Equation (\ref{brktime}), with these three constraints we calculate the region of overlap of these quantities on the $M_{\bullet}$-$a$ plane.

\begin{figure}
\centerline{\includegraphics[scale=0.3]{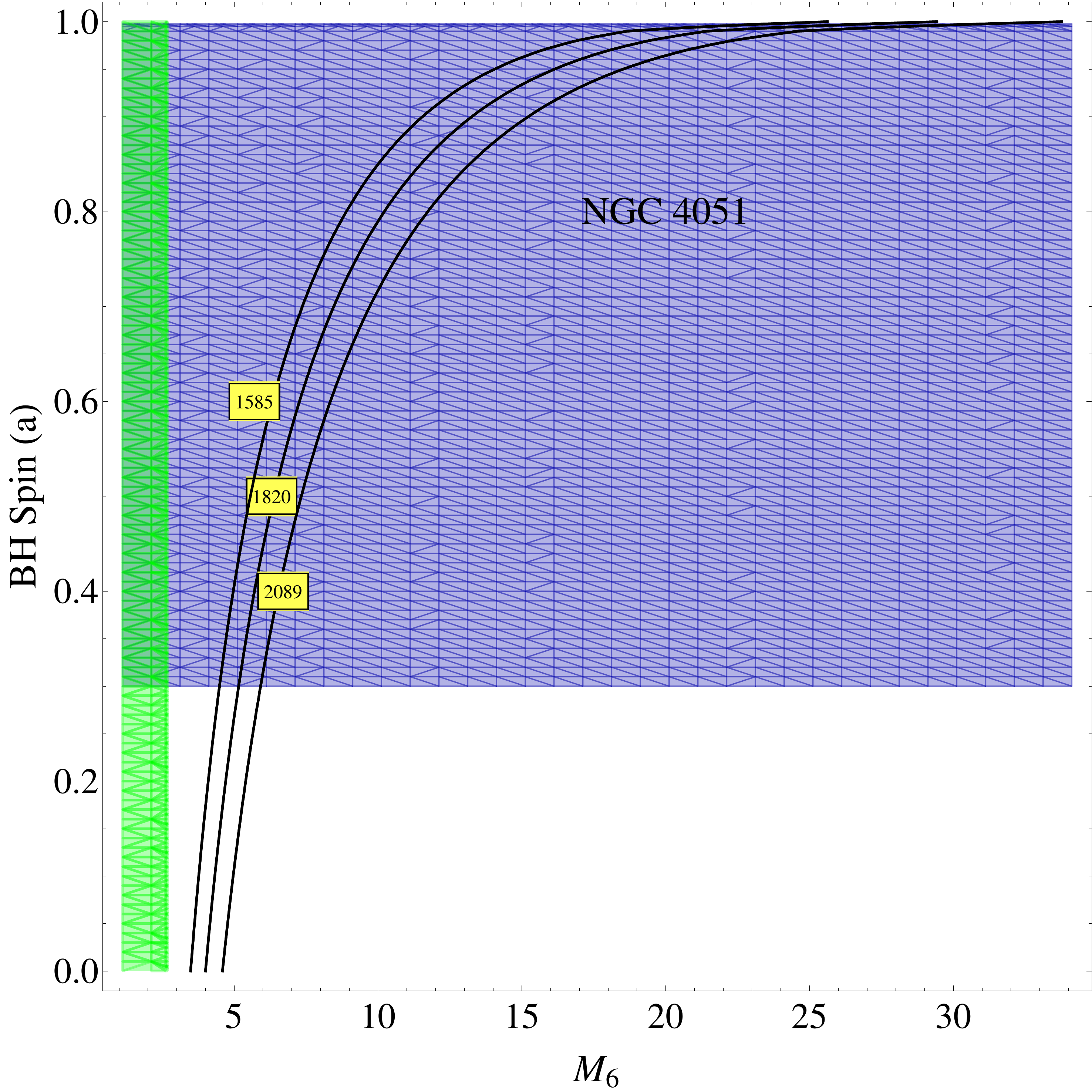}}
\caption{Contours of the break timescale $T_{B} = T_{B}(r,a)$ plotted for the NLSy1 galaxy NGC 4051 (ObsIDs 0157560101, 0606320101, 0606320201 and 0606321601). Three constraint bands are plotted in the figure. The first is from the black hole mass of $1.90^{+0.78}_{-0.78} \times 10^6 M_\odot$, plotted as a vertical green band. The second is from the black hole spin $\geq$ 0.30, plotted as a horizontal blue band. The third is from the contours of $T_{B}$, which are inferred from the current analysis to be 1820 s lying within a 68 \% confidence interval from 1585 s and 2089 s, plotted as the contour bands. Only two of the constraint bands intersect (spin and $T_{B}$). An upper limit on the black hole mass of $2.85 \times 10^7 M_\odot$ can be inferred from the current analysis.}
\label{NGC4051brkfig}
\end{figure}

NGC 4051 with an inferred break timescale of $1820^{+269}_{-235}$ (from the current analysis) hosts a BH of mass $(1.9^{+0.78}_{-0.78})\times 10^6 M_\odot$ and a lower limit on the spin of $\geq$ 0.30. Contours of $T_{B} = T_{B}(r,a)$ are plotted in Figure \ref{NGC4051brkfig}. The overlap in this case is only between the contours of $T_{B}$ and the spin $a$. An upper limit to the black hole mass of $2.85 \times 10^7 M_\odot$ (for a maximally spinning black hole with $a = 0.998$) can be inferred.

\begin{figure}
\centerline{\includegraphics[scale=0.3]{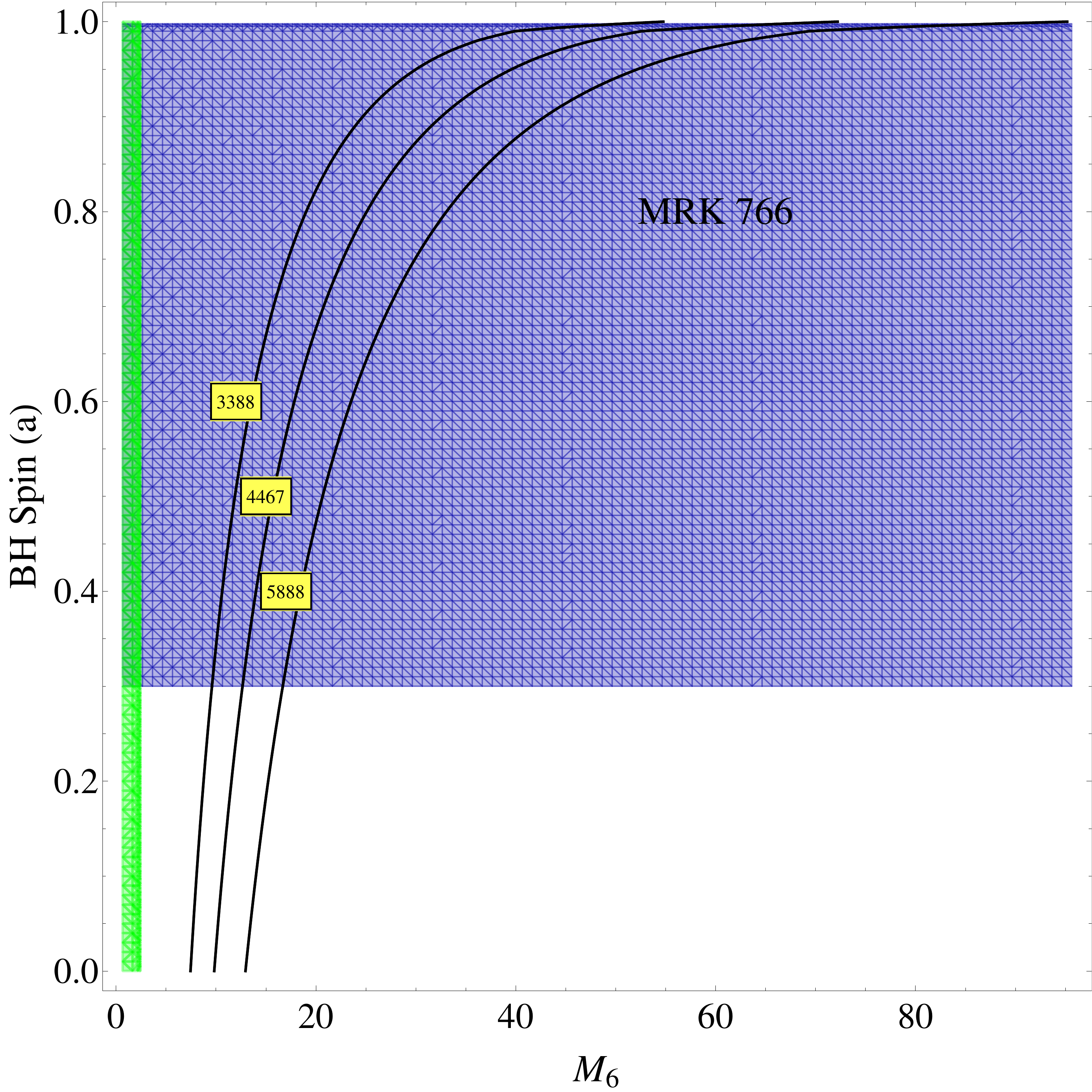}}
\caption{Contours of the break timescale $T_{B} = T_{B}(r,a)$ plotted for the NLSy1 galaxy MRK 766 (ObsIDs 0096020101, 0304030301 and 0304030601). Three constraint bands are plotted in the figure. The first is from the black hole mass of $1.26^{+1.19}_{-0.61} \times 10^6 M_\odot$, plotted as a vertical green band. The second is from the black hole spin $\geq$ 0.30, plotted as a horizontal blue band. The third is from the contours of $T_{B}$, which are inferred from the current analysis to be 4467 s lying within a 68 \% confidence interval from 3388 s and 5888 s, plotted as the contour bands. Only two of the constraint bands intersect (spin and $T_{B}$). An upper limit on the black hole mass of $8.02 \times 10^7 M_\odot$ can be inferred from the current analysis.}
\label{MRK766brkfig}
\end{figure}

MRK 766 with an inferred weighted mean break timescale of $4467^{+1421}_{-1079}$ (from the current analysis) hosts a BH of mass $(1.26^{+1.19}_{-0.61})\times 10^6 M_\odot$ and a lower limit on the spin of $\geq$ 0.30. Contours of $T_{B} = T_{B}(r,a)$ are plotted in Figure \ref{MRK766brkfig}. The overlap in this case is only between the contours of $T_{B}$ and the spin $a$. An upper limit to the black hole mass of $8.02 \times 10^7 M_\odot$ (for a maximally spinning black hole with $a = 0.998$) can be inferred.

\begin{figure}
\centerline{\includegraphics[scale=0.3]{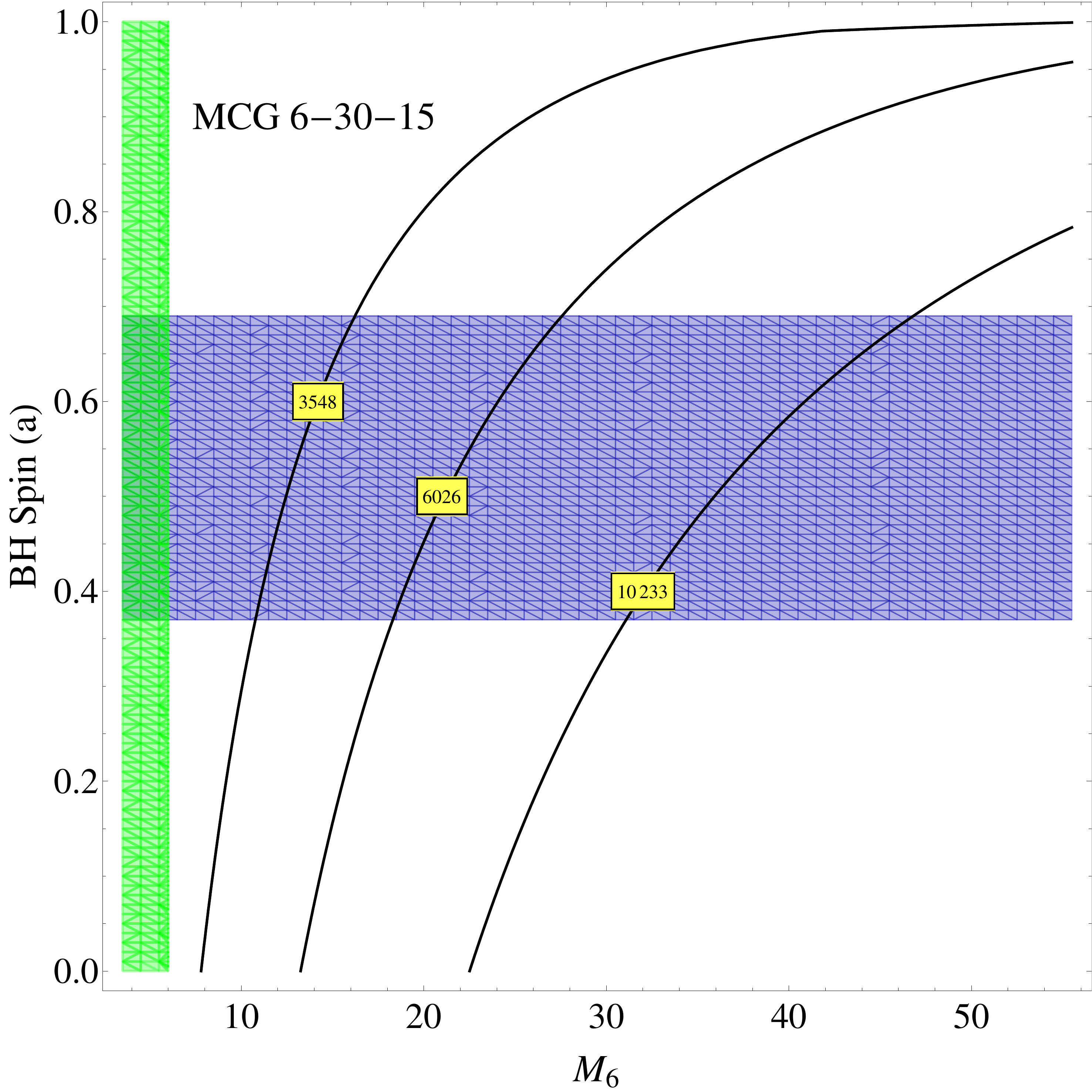}}
\caption{Contours of the break timescale $T_{B} = T_{B}(r,a)$ plotted for the Sy1 galaxy MCG-6-30-15 (ObsIDs 0029740801 and 0111570101). Three constraint bands are plotted in the figure. The first is from the black hole mass of $4.5^{+1.5}_{-1.0} \times 10^6 M_\odot$, plotted as a vertical green band. The second is from the black hole spin 0.49$^{+0.20}_{-0.12}$, plotted as a horizontal blue band. The third is from the contours of $T_{B}$, which are inferred from the current analysis to be 6026 s lying within a 68 \% confidence interval from 3548 s and 10233 s, plotted as the contour bands. Only two of the constraint bands intersect (spin and $T_{B}$). An upper limit on the black hole mass of $4.68 \times 10^7 M_\odot$ can be inferred from the current analysis.}
\label{MCGbrkfig}
\end{figure}

MCG-6-30-15 with an inferred break timescale of $6026^{+4207}_{-2478}$  (from the current analysis) hosts a BH of mass $(1.9^{+0.78}_{-0.78}) \times 10^6 M_\odot$ and a lower limit on the spin of $\geq$ 0.80. Contours of $T_{B} = T_{B}(r,a)$ are plotted in Figure \ref{MCGbrkfig}. The overlap in this case is only between the contours of $T_{B}$ and the spin $a$. An upper limit to the black hole mass of $4.68\times 10^7 M_\odot$ (for a maximally spinning black hole with $a = 0.998$) can be inferred.

The main results from our analysis which include constraints on $M_{\bullet}$ and $a$ from the above discussion are presented in Table \ref{Breaktable}.  We do not find any object for which all three constraint regions overlap. If this was the case, the BH mass and spin can be constrained simultaneously.

\section{Conclusions}
\label{conclusions}

Simple theoretical models based on orbital features of the accretion disk are proposed for observed short timescales due to X-ray variability in AGNs. These include constraints on the extent of the emission region, the dynamic timescale, an associated quality factor $Q$, and a break frequency model, all being cast in Kerr geometry. Simulations of $Q$ are used to clearly demarcate between three regions of applicability in the thin disk as $Q = 0.15 - 0.23$ in the plunging region (gas pressure dominated, opacity due to electron scattering), $Q = 2.76\times 10^3 - 4.25 \times 10^{6}$ in the edge region (at or close to $r_{ISCO}$, gas pressure dominated, opacity due to electron scattering) and $Q = 6.3 \times 10^{-2} - 3.5 \times 10^4$ in the inner region (close to $r_{ISCO}$, radiation-pressure-dominated, opacity due to electron scattering) for a range of viscosity $\alpha$ between 0.01 and 0.4. The simulated $Q$ values indicate a distinct demarcation between the plunge, edge and inner regions. The simulations also show that $Q$ is robust against changes in $\alpha$. We were able to identify the inner region as the physically relevant region as the simulated $Q$ are close to that measured from the QPO in the NLS1 galaxy REJ 1034+396. 

An analysis of X-ray light curves ($0.3-10$ keV) from a group of Seyfert galaxies is conducted to study its variability properties expressed in terms of the excess fractional variability $F_{\mathrm{var}}$ and to infer any statistically significant break timescale. The slope $\alpha$ in the relationship $F_{\mathrm{var}} \propto M^{\alpha}_{\bullet}$ is found to be $-0.33 \pm 0.12$, flatter than the expected slope of $-0.5$ for that determined from the light curves of the Sy1 galaxies. If we include data from the NLSy1s, then the slope steepens to $-0.40 \pm 0.08$, consistent within error bars of previous studies. The flatter nature of the slope could be attributed to a smaller sample size of this study.

The periodogram fit procedure we use does not resort to any numerical differentiation scheme e.g. \cite{2012ApJ...746..131B}, which may introduce possible artifacts of particular methods used in carrying it out. The fit procedure and model selection can be extended to any physically motivated PSD model. Some of the X-ray light curves from the Sy1s and NLSy1s in our study were previously analyzed in a timing study of various classes of radio-quiet as well as radio-loud AGNs, the entire sample consisting of 104 light curves from 209 nearby AGN ($z<$0.4) in order to identify and constrain the PSD shape \citep{2012A&A...544A..80G}. A power-law and a bending power-law model are used to fit the periodogram and infer the underlying PSD shape with model selection being carried out using the likelihood ratio test. These objects include MRK 335, Fairall 9, 3C 120, ARK 120, MRK 79, NGC 3516, NGC 3783, NGC 4051, NGC 4151, MRK 766, MCG-6-30-015, IC 4329A and MRK 509. Our results for many of these objects are consistent with the study where a power-law PSD shape is clearly detected. As our study is physically motivated, we restrict the power law slope to $<$ 2.5. This could lead to some differences in the quoted results.

The light curves do not indicate the presence of any statistically significant QPO. Bend timescales ranging between 1820 s and 6026 s are inferred from this analysis in nine light curves from two NLSy1 galaxies (NGC 4051, MRK 766) and one Seyfert galaxy (MCG-6-30-15). In NGC 4051, MRK 766 and MCG-6-30-15, revised upper limits on BH masses of $2.85 \times 10^7 M_\odot$, $8.02 \times 10^7 M_\odot$ and $4.68 \times 10^7 M_\odot$ are inferred. The dynamic timescale and $Q$-factor models are applied to the X-ray light curve of REJ 1034+396 where a statistically significant QPO at 3733 s was inferred from a previous study. With the measured $Q$ of 32.0 $\pm$ 6.5, an emission radius of $\sim$ 6 $M - 6.5 M$ is inferred. The BH spin $a$ is inferred to be $\leq$ 0.08 for $\dot{m}$ $\leq$ 0.3. As the $Q$ measured falls exactly within the range of simulated $Q$ for the inner region close to $r_{ISCO}$, there is strong evidence to suggest that REJ 1034+396 hosts a relativistic thin disk.

If a light curve indicates both a QPO and a break timescale, then the QPO can be used to constrain $r$, $a$ and $\dot{m}$ if the black hole mass is known using Equations (\ref{time}) and (\ref{Qfac}). The $r$ and $a$ must satisfy the inequalities Equations (\ref{eqn1}) and (\ref{eqn2}), thus acting as a consistency check as to the relevance of the theoretical models developed. The relative simplicity of the theoretical models make them easily applicable to different types of accretion disk models so they allow one to infer the black hole properties in a robust and statistically sound manner. 

\section*{Acknowledgments}
We thank the referee for suggesting changes to the manuscript which has significantly improved both its content and presentation. We thank Professor Paul J. Wiita for helpful suggestions. We also thank Sandra Rajiva for proof reading the manuscript. Figures \ref{Xrayplots1}, \ref{Xrayplots2}, and \ref{Xrayplots3} in this manuscript were created using the LevelScheme scientific figure preparation system \citep{2005CoPhC.171..107C}. This work is based on observations obtained with {\em XMM-Newton}, an ESA science mission with instruments and contributions directly funded by ESA member States and the USA (NASA).

\bibliography{main}

\appendix
\section{Calculation of Equation (2)}
\label{appA}

Equation (\ref{omega}) is applicable only for a stationary observer in the Kerr metric ($r$ and $\theta$ are fixed). If we restrict the motion to the equatorial plane ($\theta = \pi/2$), then the components of the Kerr metric are given by,

\be
g_{t \phi}=-{4 a M r \over \Sigma}
\ee
\be
g_{tt}=-\left(1-{2 M r \Sigma}\right)
\ee
\be
g_{\phi \phi}=\left(r^2+a^2+{2 M r a^2 \over \Sigma}\right) 
\ee
where $\Sigma = r^2+a^2 \cos^2 \theta = r^2$.
These can be substituted into Equation (\ref{omega}), which then gives
\be
\disp \Omega_{\rm max, min} ={{4 a M \over r} \pm \left({16 a^2 M^2 \over r^2}+\left(1-{2 M \over r}\right) \left(r^2+a^2+{2 M a^2 \over r}\right)\right)^{1/2} \over {\left(r^2+a^2+{2 M a^2 \over r}\right)}}.
\ee
On making the substitution $r/M \rightarrow r$ and $a/M \rightarrow a$ and simplifying $\Omega_{\rm min} < \Omega_K <\Omega_{\rm max}$, we obtain
\be
{{4 a \over r} - \left({12 a^2 \over r^2}+a^2+r^2-2 r\right)^{1/2} \over {\left(r^2+a^2+{2 a^2 \over r}\right)}} < {1 \over {r^{3/2}+a}} < {{4 a \over r} + \left({12 a^2 \over r^2}+a^2+r^2-2 r\right)^{1/2} \over {\left(r^2+a^2+{2 a^2 \over r}\right)}}
\ee
Simplifying inequality by rearranging the terms and squaring, we finally obtain Equation (\ref{eqn1}).


\begin{table}[!h]
\centerline{{\bf Summary of AGN properties}}
%\caption[Reported properties of sample of Seyfert galaxies]
\centerline{
%.8
\scalebox{.9}{
\begin{tabular}{|l|l|l|l|l|l|l|l|}
\hline
Object & AGN     & RA      & Dec.    & red-shift & Black hole                  & Black      & Average \\
       & Type    &         &         & z        & mass                        & hole       & Fractional \\
       &         &         &         &          & $M_{\bullet}$                     & spin       & Variability \\
       &         &         &         &          & ($10^6 M_\odot$); ref.   & $a$; ref.     & $F_{\mathrm{var}}$ \\ \hline
       &         &         &         &          &                             &           &                \\
MRK 335     & NLSy1   & 00$^{\mathrm{h}}$06$^{\mathrm{m}}$19.5$^{\mathrm{s}}$  & +20$^{\mathrm{d}}$12$^{\mathrm{m}}$10$^{\mathrm{s}}$  & 0.258 & (14.2$^{+3.7}_{-3.7}$); 1      & (0.83$^{+0.09}_{-0.13}$); 7 & 18.66   \\
Q 0056-363  & Sy1     & 00$^{\mathrm{h}}$58$^{\mathrm{m}}$37.3$^{\mathrm{s}}$  & -36$^{\mathrm{d}}$06$^{\mathrm{m}}$05$^{\mathrm{s}}$  & 0.164 & 610; 2          & NA           & 7.85   \\
Fairall 9   & Sy1.2   & 01$^{\mathrm{h}}$23$^{\mathrm{m}}$45.8$^{\mathrm{s}}$  & -58$^{\mathrm{d}}$48$^{\mathrm{m}}$21$^{\mathrm{s}}$  & 0.047 & (255$^{+56}_{-56}$); 1 & (0.82$^{+0.09}_{-0.19}$); 7          & 6.53 \\
3C 120      & Sy1.5   & 04$^{\mathrm{h}}$33$^{\mathrm{m}}$11.1$^{\mathrm{s}}$  & +05$^{\mathrm{d}}$21$^{\mathrm{m}}$16$^{\mathrm{s}}$  & 0.033 & (55.5$^{+31.4}_{-22.5}$); 1 & NA           &  5.97  \\
ARK 120     & Sy1     & 05$^{\mathrm{h}}$16$^{\mathrm{m}}$11.4$^{\mathrm{s}}$  & -00$^{\mathrm{d}}$08$^{\mathrm{m}}$59$^{\mathrm{s}}$  & 0.033 & (150$^{+19}_{-19}$); 1 & (0.64$^{+0.19}_{-0.11}$); 7          & 2.91\\
MRK 79      & Sy1.2   & 07$^{\mathrm{h}}$42$^{\mathrm{m}}$32.8$^{\mathrm{s}}$  & +49$^{\mathrm{d}}$48$^{\mathrm{m}}$35$^{\mathrm{s}}$  & 0.022 & (52.4$^{+14.4}_{-14.4}$); 1  &  (0.7$^{+0.1}_{-0.1}$); 8     & 8.87\\
MCG-5-23-16 & Sy1     & 09$^{\mathrm{h}}$47$^{\mathrm{m}}$40.1$^{\mathrm{s}}$  & -30$^{\mathrm{d}}$56$^{\mathrm{m}}$55$^{\mathrm{s}}$  & 0.008 & (83.2$^{+125.7}_{-50.1}$); 3          & NA           & 8.09   \\
NGC 3516    & Sy1.5   & 11$^{\mathrm{h}}$06$^{\mathrm{m}}$47.5$^{\mathrm{s}}$  & +72$^{\mathrm{d}}$34$^{\mathrm{m}}$07$^{\mathrm{s}}$  & 0.009 & (42.7$^{+14.6}_{-14.6}$); 1 & $\geq$ 0.30; 9          & 14.69    \\
NGC 3783    & Sy1.5   & 11$^{\mathrm{h}}$39$^{\mathrm{m}}$01.7$^{\mathrm{s}}$  & -37$^{\mathrm{d}}$44$^{\mathrm{m}}$19$^{\mathrm{s}}$  & 0.010 & (29.8$^{+5.4}_{-5.4}$); 1       & $\geq$ 0.20; 9   & 9.26    \\
NGC 4051    & NLSy1   & 12$^{\mathrm{h}}$03$^{\mathrm{m}}$09.6$^{\mathrm{s}}$  & +44$^{\mathrm{d}}$31$^{\mathrm{m}}$53$^{\mathrm{s}}$  & 0.002 & (1.9$^{+0.78}_{-0.78}$); 1          &  $\geq$ 0.30; 9          & 39.27   \\
NGC 4151    & Sy1.5   & 12$^{\mathrm{h}}$10$^{\mathrm{m}}$32.6$^{\mathrm{s}}$  & +39$^{\mathrm{d}}$24$^{\mathrm{m}}$21$^{\mathrm{s}}$  & 0.003 & (13.3$^{+4.6}_{-4.6}$); 1          &  NA          &  6.41   \\
MRK 766     & NLSy1   & 12$^{\mathrm{h}}$18$^{\mathrm{m}}$26.5$^{\mathrm{s}}$  & +29$^{\mathrm{d}}$48$^{\mathrm{m}}$46$^{\mathrm{s}}$  & 0.013 & (1.26$^{+1.19}_{-0.61}$); 4          & $\geq$ 0.30; 9 & 25.82    \\
MCG-6-30-15 & Sy1.2   & 13$^{\mathrm{h}}$35$^{\mathrm{m}}$53.7$^{\mathrm{s}}$  & -34$^{\mathrm{d}}$17$^{\mathrm{m}}$44$^{\mathrm{s}}$  & 0.008 & (4.5$^{+1.5}_{-1.0}$); 5        & 0.49$^{+0.20}_{-0.12}$; 10 & 26.89   \\
IC 4329A    & Sy1.2   & 13$^{\mathrm{h}}$49$^{\mathrm{m}}$19.2$^{\mathrm{s}}$  & -30$^{\mathrm{d}}$18$^{\mathrm{m}}$34$^{\mathrm{s}}$  & 0.016 & (218.88$^{+217.64}_{-109.65}$); 6 & $\geq$ 0.00; 9        & 3.76\\
MRK 509     & Sy1.5   & 20$^{\mathrm{h}}$44$^{\mathrm{m}}$09.7$^{\mathrm{s}}$  & -10$^{\mathrm{d}}$43$^{\mathrm{m}}$25$^{\mathrm{s}}$  & 0.034 & (143$^{+12}_{-12}$); 1 & (0.78$^{+0.03}_{-0.04}$); 7 & 3.34 \\
NGC 7469    & Sy1.5   & 23$^{\mathrm{h}}$03$^{\mathrm{m}}$15.6$^{\mathrm{s}}$  & +08$^{\mathrm{d}}$52$^{\mathrm{m}}$26$^{\mathrm{s}}$  & 0.016 & (12.2$^{+1.4}_{-1.4}$); 1         & (0.64$^{+0.13}_{-0.20}$); 7  & 5.55 \\ \hline
\end{tabular}}} %\vspace{.1cm}
\caption{BH Mass: $^1$: \cite{2004ApJ...613..682P} (virial mass using reverberation mapping with mass calibrated using the $M_{\bullet}$-$\sigma$ relation); $^2$: \cite{2003A&A...408..119P} ;$^3$: \cite{1995A&A...301...55O} (M-$\sigma$ relation); $^4$: \cite{2013MNRAS.431.2441D}; $^5$: \cite{2005MNRAS.359.1469M} (use of the scaling of the PSD break timescale with black hole mass and confirmation with reverberation mapping);  $^6$: \cite{2009ApJ...698.1740M}  (M-$\sigma$ relation); BH spin: $^7$: \cite{2013MNRAS.428.2901W}; $^8$: \cite{2011MNRAS.411..607G}; $^9$: \cite{2010A&A...524A..50D}; $^{10}$: \cite{2011MNRAS.416.2725P}; NA: not available. Columns 1 - 8 give the object name, its right ascension (RA), declination (Dec.), cosmological red-shift $z$, the mass of the SMBH it hosts $M_{\bullet}$ in terms of $10^6 M_\odot$, the spin of the black hole and the average excess fractional variability index $F_{\mathrm{var}}$ determined in the current study.}
\label{AGNsummary}
\end{table}
\pagebreak
\newpage

\centerline{{\bf PSD fit results}}
\begin{center}
\scriptsize{
\begin{longtable}{|l|l|l|l|l|l|l|l|l|l|}
\hline
Object   & Observation & Time     & Fractional        & PSD       & \multicolumn{3}{|l|}{PSD Fit parameters}    & AIC & Model \\
         & ID          & Duration & Variability       & model     & \multicolumn{3}{|l|}{} &     & likelihood\\
         &             & (s)      & $F_{\mathrm{var}}$&           & log(A) & $\alpha$ & log(f$_b$)            &     & \\ \hline
MRK 335 & 0101040101 & 31500 & 11.55 & PL & -6.1 $\pm$ 0.4 & -1.9 $\pm$ 0.2 & & 333.60 & 1.00 \\
 & & & & BPL& -2.5 $\pm$ 0.2 & -2.3 $\pm$ 0.2 & -3.60 $\pm$ 0.28 & 333.92 & 0.85 \\
 & 0306870101 & 132100 & 12.18 & {\bf PL} & -6.1 $\pm$ 0.2 & -1.9 $\pm$ 0.1 & & 1379.93 & 1.00 \\
 & & & & BPL& -2.1 $\pm$ 0.3 & -2.0 $\pm$ 0.1 & -4.22 $\pm$ 0.41 & 1386.12 & 0.05 \\
 & 0600540501 & 80600 & 24.17 & {\bf PL} & -5.0 $\pm$ 0.2 & -1.7 $\pm$ 0.1 & & 90.33 & 1.00 \\
 & & & & BPL& -1.9 $\pm$ 0.1 & -1.9 $\pm$ 0.2 & -4.00 $\pm$ 0.32 & 98.73 & 0.02 \\
 & 0600540601 & 111600 & 26.75 & PL & -4.6 $\pm$ 0.2 & -1.6 $\pm$ 0.1 & & 590.83 & 0.07 \\
 & & & & BPL& -2.5 $\pm$ 0.1 & -3.5 $\pm$ 0.2 & -2.72 $\pm$ 0.03 & 583.75 & 1.00 \\
Q 0056-363 & 0205680101 & 101000 & 7.77 & PL & -3.4 $\pm$ 0.1 & -1.1 $\pm$ 0.1 & & 78.97 & 10$^{-11}$ \\
 & & & & BPL& -2.7 $\pm$ 0.1 & -3.5 $\pm$ 0.1 & -2.70 $\pm$ 0.02 & 29.09 & 1.00 \\
 & 0401930101 & 44600 & 7.92 & PL & -3.1 $\pm$ 0.4 & -1.0 $\pm$ 0.2 & & 118.26 & 0.04 \\
 & & & & BPL& -2.5 $\pm$ 0.1 & -3.2 $\pm$ 0.3 & -2.71 $\pm$ 0.06 & 111.78 & 1.00 \\
Fairall 9 & 0101040201 & 28800 & 4.27 & {\bf PL} & -3.9 $\pm$ 0.4 & -1.1 $\pm$ 0.1 & & 307.55 & 1.00 \\
 & & & & BPL& -3.3 $\pm$ 0.1 & -3.5 $\pm$ 0.2 & -2.63 $\pm$ 0.05 & 329.59 & 10$^{-5}$ \\
 & 0605800401 & 129400 & 8.79 & {\bf PL} & -4.9 $\pm$ 0.1 & -1.5 $\pm$ 0.1 & & 1111.29 & 1.00 \\
 & & & & BPL& -3.0 $\pm$ 0.1 & -3.5 $\pm$ 0.1 & -2.73 $\pm$ 0.02 & 1182.69 & 10$^{-16}$ \\  
3C 120 & 0152840101 & 125400 & 5.97 & {\bf PL} & -4.7 $\pm$ 0.1 & -1.3 $\pm$ 0.1 & & 1925.5 & 1.00 \\
 & & & & BPL& -3.3 $\pm$ 0.1 & -3.5 $\pm$ 0.2 & -2.72 $\pm$ 0.03 & 1956.96 & 10$^{-7}$ \\
ARK 120 & 0147190101 & 111100 & 2.91 & {\bf PL} & -4.7 $\pm$ 0.1 & -1.2 $\pm$ 0.1 & & 2332.55 & 1.00 \\
 & & & & BPL& -3.6 $\pm$ 0.1 & -3.5 $\pm$ 0.1 & -2.69 $\pm$ 0.02 & 2401.49 & 10$^{-15}$ \\ 
MRK 79 & 0502091001 & 78100 & 8.87 & {\bf PL} & -3.3 $\pm$ 0.2 & -1.1 $\pm$ 0.1 & & 239.16 & 1.00 \\
 & & & & BPL& -2.5 $\pm$ 0.1 & -1.3 $\pm$ 0.1 & -3.99 $\pm$ 0.37 & 246.02 & 0.03 \\   
MCG-5-23-16 & 0302850201 & 116800 & 8.09 & {\bf PL} & -5.5 $\pm$ 0.2 & -1.6 $\pm$ 0.1 & & 1920.70 & 1.00 \\
 & & & & BPL& -3.4 $\pm$ 0.1 & -3.5 $\pm$ 0.1 & -2.72 $\pm$ 0.02 & 1940.36 & 10$^{-4}$ \\
NGC 3516 & 0107460601 & 125900 & 18.35 & {\bf PL} & -5.0 $\pm$ 0.2 & -1.6 $\pm$ 0.1 & & 424.95 & 1.00 \\
 & & & & BPL& -2.8 $\pm$ 0.1 & -3.5 $\pm$ 0.1 & -2.73 $\pm$ 0.02 & 461.52 & 10$^{-8}$ \\ 
 & 0107460701 & 127800 & 8.88 & {\bf PL} & -3.5 $\pm$ 0.1 & -1.1 $\pm$ 0.1 & & 57.29 & 1.00 \\
 & & & & BPL& -2.7 $\pm$ 0.1 & -3.5 $\pm$ 0.1 & -2.70 $\pm$ 0.02 & 100.67 & 10$^{-10}$ \\
 & 0401210401 & 51500 & 14.95 & PL & -5.5 $\pm$ 0.2 & -1.7 $\pm$ 0.1 & & 518.80 & 0.03 \\
 & & & & BPL& -2.3 $\pm$ 0.2 & -1.8 $\pm$ 0.2 & -4.24 $\pm$ 0.25 & 511.79 & 1.00 \\
 & 0401210501 & 68400 & 8.37 & PL & -5.5 $\pm$ 0.2 & -1.6 $\pm$ 0.1 & & 1035.55 & 1.00 \\
 & & & & BPL& -3.4 $\pm$ 0.1 & -3.5 $\pm$ 0.3 & -2.68 $\pm$ 0.27 & 1035.74 & 0.91 \\
 & 0401210601 & 67900 & 28.25 & PL & -6.1 $\pm$ 0.3 & -2.0 $\pm$ 0.1 & & 295.61 & 1.00 \\
 & & & & BPL& -1.9 $\pm$ 0.3 & -2.2 $\pm$ 0.2 & -3.93 $\pm$ 0.38 & 296.83 & 0.54 \\
 & 0401211001 & 59300 & 9.36 & PL & -5.7 $\pm$ 0.3 & -1.7 $\pm$ 0.1 & & 848.18 & 0.12 \\
 & & & & BPL& -3.4 $\pm$ 0.3 & -3.5 $\pm$ 0.5 & -2.72 $\pm$ 0.5 & 844.02 & 1.00 \\
NGC 3783 & 0112210101 & 37000 & 6.90 & {\bf PL} & -5.1 $\pm$ 0.4 & -1.5 $\pm$ 0.1 & & 407.08 & 1.00 \\
 & & & & BPL& -3.26 $\pm$ 0.1 & -3.5 $\pm$ 0.5 & -2.65 $\pm$ 0.08 & 414.22 & 0.03 \\
 & 0112210201 & 57100 & 8.19 & {\bf PL} & -4.6 $\pm$ 0.3 & -1.4 $\pm$ 0.1 & & 498.94 & 1.00 \\
 & & & & BPL& -3.1 $\pm$ 0.1 & -3.5 $\pm$ 0.6 & -2.67 $\pm$ 0.15 & 503.68 & 0.09 \\
 & 0112210501 & 136300 & 12.68 & {\bf PL} & -5.1 $\pm$ 0.2 & -1.5 $\pm$ 0.1 & & 1701.80 & 1.00 \\
 & & & & BPL& -3.2 $\pm$ 0.1 & -3.5 $\pm$ 0.1 & -2.72 $\pm$ 0.02 & 1724.44 & 10$^{-5}$ \\
NGC 4051 & 0157560101 & 33600 & 21.59 & PL & -5.7 $\pm$ 0.5 & -2.0 $\pm$ 0.2 & & 178.81 & 0.05 \\
 & & & & {\bf BPL}& -1.7 $\pm$ 0.2 & -2.7 $\pm$ 0.4 & -3.41 $\pm$ 0.33 & 171.08 & 1.00 \\
 & 0606320101 & 45200 & 34.77 & PL & -6.6 $\pm$ 0.4 & -2.4 $\pm$ 0.2 & & 415.99 & 10$^{-7}$ \\
 & & & & {\bf BPL} & -1.6 $\pm$ 0.1 & -3.3 $\pm$ 0.2 & -3.21 $\pm$ 0.09 & 384.79 & 1.00 \\
 & 0606320201 & 43400 & 47.22 & PL & -6.5 $\pm$ 0.3 & -2.5 $\pm$ 0.1 & & 722.33 & 10$^{-5}$ \\
 & & & & {\bf BPL}& -1.1 $\pm$ 0.2 & -2.9 $\pm$ 0.2 & -3.38 $\pm$ 0.16 & 699.74 & 1.00 \\
 & 0606321601 & 41300 & 53.48 & PL & -7.0 $\pm$ 0.4 & -2.6 $\pm$ 0.1 & & 557.00 & 10$^{-7}$ \\
 & & & & {\bf BPL}& -1.3 $\pm$ 0.2 & -3.2 $\pm$ 0.2 & -3.26 $\pm$ 0.12 & 527.28 & 1.00 \\ 
\hline
\newpage
%\centerline{PSD fit results: continued from previous page}
\hline
Object   & Observation & Time     & Fractional        & PSD       & \multicolumn{3}{|l|}{PSD Fit parameters}    & AIC & Model \\
         & ID          & Duration & Variability       & model     & \multicolumn{3}{|l|}{} &     & likelihood\\
         &             & (s)      & $F_{\mathrm{var}}$&           & log(A) & $\alpha$ & log(f$_b$)            &     & \\
\hline
NGC 4151 & 0402660101 & 39700 & 5.85 & {\bf PL} & -3.6 $\pm$ 0.4 & -1.1 $\pm$ 0.2 & & 122.00 & 1.00 \\
 & & & & BPL& -2.9 $\pm$ 0.1 & -3.5 $\pm$ 0.1 & -2.71 $\pm$ 0.04 & 161.54 & 10$^{-9}$ \\
 & 0402660201 & 47300 & 6.97 & {\bf PL} & -5.3 $\pm$ 0.4 & -1.6 $\pm$ 0.1 & & 501.18 & 1.00 \\
 & & & & BPL& -3.1 $\pm$ 0.1 & -3.4 $\pm$ 0.4 & -2.76 $\pm$ 0.09 & 513.38 & 10$^{-3}$ \\
MRK 766 & 0096020101 & 38800 & 19.22 & PL & -5.4 $\pm$ 0.4 & -1.8 $\pm$ 0.1 & & 33.91 & 0.08 \\
 & & & & {\bf BPL} & -1.7 $\pm$ 0.2 & -2.1 $\pm$ 0.2 & -4.11 $\pm$ 0.21 & 28.86 & 1.00 \\
 & 0109141301 & 128400 & 26.29 & {\bf PL} & -6.2 $\pm$ 0.2 & -2.1 $\pm$ 0.1 & & 91.57 & 1.00 \\
 & & & & BPL& -2.1 $\pm$ 0.2 & -2.6 $\pm$ 0.2 & -3.38 $\pm$ 0.25 & 115.80 & 10$^{-5}$ \\
 & 0304030301 & 98300 & 37.90 & PL & -6.0 $\pm$ 0.2 & -2.1 $\pm$ 0.1 & & 418.47 & 0.01 \\
 & & & & {\bf BPL} & -1.6 $\pm$ 0.4 & -2.4 $\pm$ 0.2 & -3.74 $\pm$ 0.42 & 408.37 & 1.00 \\
 & 0304030401 & 94300 & 22.49 & {\bf PL} & -6.5 $\pm$ 0.3 & -2.2 $\pm$ 0.1 & & 217.53 & 1.00 \\
 & & & & BPL& -2.1 $\pm$ 0.1 & -2.7 $\pm$ 0.2 & -3.34 $\pm$ 0.16 & 253.43 & 10$^{-8}$ \\
 & 0304030501 & 94100 & 19.97 & {\bf PL} & -7.1 $\pm$ 0.3 & -2.4 $\pm$ 0.1 & & 332.28 & 1.00 \\
 & & & & BPL& -2.2 $\pm$ 0.1 & -3.2 $\pm$ 0.3 & -3.17 $\pm$ 0.15 & 385.24 & 10$^{-12}$ \\
 & 0304030601 & 98300 & 28.60 & PL & -6.0 $\pm$ 0.2 & -2.1 $\pm$ 0.1 & & 463.58 & 10$^{-7}$ \\
 & & & & {\bf BPL}& -1.79 $\pm$ 0.1 & -2.6 $\pm$ 0.2 & -3.41 $\pm$ 0.15 & 433.89 & 1.00 \\
 & 0304030701 & 29000 & 26.29 & PL & -6.0 $\pm$ 0.3 & -2.1 $\pm$ 0.1 & & 83.00 & 1.00 \\
 & & & & BPL& -1.4 $\pm$ 0.3 & -2.3 $\pm$ 0.2 & -3.99 $\pm$ 0.26 & 84.16 & 0.56 \\
MCG-6-30-15 & 0029740101 & 80500 & 29.25 & {\bf PL} & -7.5 $\pm$ 0.3 & -2.5 $\pm$ 0.1 & & 453.53 & 1.00 \\
 & & & & BPL& -1.8 $\pm$ 0.2 & -2.9 $\pm$ 0.2 & -3.54 $\pm$ 0.17 & 479.95 & 10$^{-6}$ \\
 & 0029740701 & 12300 & 19.78 & {\bf PL} & -7.0 $\pm$ 0.2 & -2.3 $\pm$ 0.1 & & 774.50 & 1.00 \\
 & & & & BPL& -2.1 $\pm$ 0.1 & -2.7 $\pm$ 0.2 & -3.51 $\pm$ 0.13 & 807.67 & 10$^{-7}$ \\
 & 0029740801 & 124100 & 35.67 & PL & -6.4 $\pm$ 0.2 & -2.2 $\pm$ 0.1 & & 221.35 & 10$^{-4}$ \\
 & & & & {\bf BPL} & -1.4 $\pm$ 0.3 & -2.3 $\pm$ 0.2 & -3.98 $\pm$ 0.30 & 203.57 & 1.00 \\
 & 0111570101 & 43100 & 30.16 & PL & -6.3 $\pm$ 0.3 & -2.2 $\pm$ 0.1 & & 85.69 & 0.11 \\
        &            &       &       & {\bf BPL}& -1.7 $\pm$ 0.3 & -2.6 $\pm$ 0.3 & -3.52 $\pm$ 0.35 & 81.20 & 1.00 \\
 & 0111570201 & 53400 & 19.58 & {\bf PL} & -7.5 $\pm$ 0.3 & -2.5 $\pm$ 0.1 & & 330.82 & 1.00 \\
        &            &       &       & BPL& -2.0 $\pm$ 0.2 & -3.1 $\pm$ 0.2 & -3.37 $\pm$ 0.16 & 358.62 & 10$^{-6}$ \\
IC 4329A & 0147440101 & 132400 & 3.76 & {\bf PL} & -4.9 $\pm$ 0.2 & -1.3 $\pm$ 0.1 & & 2744.69 & 1.00 \\
 & & & & BPL& -3.6 $\pm$ 0.1 & -3.5 $\pm$ 0.2 & -2.70 $\pm$ 0.03 & 2778.72 & 10$^{-8}$ \\ 
MRK 509 & 0130720101 & 29200 & 3.15 & {\bf PL} & -4.5 $\pm$ 0.5 & -1.2 $\pm$ 0.2 & & 518.01 & 1.00 \\
 & & & & BPL& -3.4 $\pm$ 0.1 & -3.5 $\pm$ 0.6 & -2.74 $\pm$ 0.23 & 526.65 & 0.01 \\ 
 & 0130720201 & 41500 & 2.55 & {\bf PL} & -4.4 $\pm$ 0.2 & -1.1 $\pm$ 0.1 & & 880.24 & 1.00 \\
 & & & & BPL& -3.6 $\pm$ 0.1 & -3.5 $\pm$ 0.7 & -2.67 $\pm$ 0.29 & 880.70 & 10$^{-4}$ \\ 
 & 0306090201 & 85200 & 2.85 & {\bf PL} & -4.4 $\pm$ 0.2 & -1.1 $\pm$ 0.1 & & 1891.38 & 1.00 \\
 & & & & BPL& -3.6 $\pm$ 0.1 & -3.5 $\pm$ 0.8 & -2.69 $\pm$ 0.39 & 1891.96 & 0.75 \\  
 & 0306090301 & 46200 & 2.41 & {\bf PL} & -4.7 $\pm$ 0.4 & -1.2 $\pm$ 0.1 & & 1038.81 & 1.00 \\
 & & & & BPL& -3.7 $\pm$ 0.1 & -3.5 $\pm$ 0.2 & -2.72 $\pm$ 0.04 & 1074.46 & 10$^{-8}$ \\ 
 & 0306090401 & 69200 & 4.37 & {\bf PL} & -4.9 $\pm$ 0.3 & -1.3 $\pm$ 0.1 & & 1456.25 & 1.00 \\
 & & & & BPL& -3.6 $\pm$ 0.1 & -3.5 $\pm$ 0.2 & -2.70 $\pm$ 0.03 & 1498.60 & 10$^{-10}$ \\ 
 & 0601390201 & 56900 & 2.87 & PL & -4.4 $\pm$ 0.3 & -1.1 $\pm$ 0.1 & & 1239.11 & 0.92 \\
 & & & & BPL& -3.6 $\pm$ 0.1 & -1.3 $\pm$ 0.7 & -3.85 $\pm$ 0.29 & 1238.96 & 1.00 \\ 
 & 0601390301 & 63100 & 2.99 & {\bf PL} & -4.5 $\pm$ 0.3 & -1.1 $\pm$ 0.1 & & 1525.85 & 1.00 \\
 & & & & BPL& -3.7 $\pm$ 0.1 & -3.5 $\pm$ 0.7 & -2.69 $\pm$ 0.28 & 1529.14 & 10$^{-3}$ \\  
 & 0601390401 & 60200 & 2.30 & {\bf PL} & -5.1 $\pm$ 0.4 & -1.3 $\pm$ 0.1 & & 1494.13 & 1.00 \\
 & & & & BPL& -3.8 $\pm$ 0.1 & -3.5 $\pm$ 0.2 & -2.73 $\pm$ 0.03 & 1545.35 & 10$^{-12}$ \\ 
 & 0601390501 & 60200 & 5.67& {\bf PL} & -5.3 $\pm$ 0.2 & -1.4 $\pm$ 0.1 & & 1407.49 & 1.00 \\
 & & & & BPL& -3.8 $\pm$ 0.1 & -3.5 $\pm$ 0.2 & -2.67 $\pm$ 0.04 & 1424.43 & 10$^{-4}$ \\ 
 & 0601390601 & 62100 & 3.48 & {\bf PL} & -5.1 $\pm$ 0.3 & -1.3 $\pm$ 0.1 & & 1544.10 & 1.00 \\
 & & & & BPL& -3.8 $\pm$ 0.1 & -3.5 $\pm$ 0.3 & -2.70 $\pm$ 0.04 & 1561.27 & 10$^{-5}$ \\ 
 & 0601390701 & 62300 & 3.60 & PL & -5.0 $\pm$ 0.3 & -1.3 $\pm$ 0.1 & & 1459.22 & 0.55 \\
 & & & & BPL& -3.7 $\pm$ 0.1 & -1.5 $\pm$ 0.2 & -3.43 $\pm$ 0.30 & 1458.05 & 1.00 \\ 

\hline
\newpage
%\centerline{PSD fit results: continued from previous page}
\hline
Object   & Observation & Time     & Fractional        & PSD       & \multicolumn{3}{|l|}{PSD Fit parameters}    & AIC & Model \\
         & ID          & Duration & Variability       & model     & \multicolumn{3}{|l|}{} &     & likelihood\\
         &             & (s)      & $F_{\mathrm{var}}$&           & log(A) & $\alpha$ & log(f$_b$)            &     & \\
\hline 
 & 0601390801 & 60200 & 4.53 & {\bf PL} & -5.0 $\pm$ 0.2 & -1.3 $\pm$ 0.1 & & 1354.61 & 1.00 \\
 & & & & BPL& -3.7 $\pm$ 0.1 & -3.5 $\pm$ 0.2 & -2.71 $\pm$ 0.04 & 1375.97 & 10$^{-5}$ \\ 
 & 0601390901 & 60200 & 2.61 & {\bf PL} & -4.8 $\pm$ 0.3 & -1.2 $\pm$ 0.1 & & 1480.51 & 1.00 \\
 & & & & BPL& -3.7 $\pm$ 0.1 & -3.5 $\pm$ 0.2 & -2.71 $\pm$ 0.04 & 1507.96 & 10$^{-6}$ \\ 
 & 0601391001 & 64600 & 3.60 & {\bf PL} & -5.1 $\pm$ 0.3 & -1.3 $\pm$ 0.1 & & 1616.49 & 1.00 \\
 & & & & BPL& -3.7 $\pm$ 0.1 & -3.5 $\pm$ 0.5 & -2.72 $\pm$ 0.14 & 1627.56 & 10$^{-3}$ \\ 
 & 0601391101 & 62100 & 3.18 & {\bf PL} & -4.8 $\pm$ 0.2 & -1.2 $\pm$ 0.1 & & 1534.24 & 1.00 \\
 & & & & BPL& -3.8 $\pm$ 0.1 & -3.5 $\pm$ 0.7 & -2.69 $\pm$ 0.28 & 1537.39 & 0.21 \\ 
NGC 7469 & 0112170301 & 22900 & 5.28 & {\bf PL} & -5.0 $\pm$ 0.4 & -1.4 $\pm$ 0.1 & & 379.75 & 1.00 \\
 & & & & BPL& -3.5 $\pm$ 0.1 & -3.5 $\pm$ 0.4 & -2.68 $\pm$ 0.07 & 389.32 & .01 \\ 
 & 0207090101 & 84300 & 5.44 & {\bf PL} & -5.1 $\pm$ 0.2 & -1.4 $\pm$ 0.1 & & 1573.14 & 1.00 \\
 & & & & BPL& -3.5 $\pm$ 0.1 & -3.5 $\pm$ 0.2 & -2.69 $\pm$ 0.03 & 1600.60 & 10$^{-6}$ \\ 
 & 0207090201 & 78400 & 5.94 & {\bf PL} & -5.3 $\pm$ 0.2 & -1.5 $\pm$ 0.1 & & 1364.64 & 1.00 \\
 & & & & BPL& -3.5 $\pm$ 0.1 & -3.5 $\pm$ 0.2 & -2.70 $\pm$ 0.04 & 1390.05 & 10$^{-6}$ \\  \hline
%\label{tab2ch5} 
%\vspace{.5cm}
\caption{}
\label{pgrambestfit}
\end{longtable}
}
\end{center} \vspace{-2.1cm}
Results from the parametric PSD models fit to the periodogram. Columns 1 -- 10 give the object name, observation I.D. from XMM Newton archives, observation duration, $F_{\mathrm{var}}$, the model (PL: power law + constant noise, BPL: bending power law + constant noise), the best-fit parameters $\log(N)$, slope $\alpha$ and the bend frequency $f_b$ with their 95\% errors derived from $\Delta S$, the AIC and the likelihood of a particular model. The best fit PSD is highlighted.
%\end{landscape}

\begin{table}[h!]
\centerline{{\bf Break frequency model applied to Seyfert light curves}}
\label{Breaktable}
\centerline{
\scalebox{1.15}{
\begin{tabular}{|l|l|l|l|}
\hline
Object & Break     & Mass   & Spin \\
       & timescale & limits & limits \\
       & $T_B$ (s) & $m_6$  & $a$ \\ \hline
& & & \\
NGC 4051 & 1820$^{+269}_{-235}$ & $\leq$ 28.5 & ($\geq$ 0.30)*  \\
& & & \\
MRK 766 & 4467$^{+1421}_{-1079}$ & $\leq$ 80.2 & ($\geq$ 0.30)* \\
& & & \\
MCG-6-30-15 & 6026$^{+4207}_{-2478}$ & $\leq$ 46.8 & (0.49$^{+0.20}_{-0.12}$)* \\
& & & \\ \hline
\end{tabular}}} %\vspace{.5cm}
\caption{Novel results from the application of the theoretical break frequency model to the X-ray light curves of Seyfert galaxies. Properties extracted include upper limits on BH mass and spin using statistically inferred $T_B$. Symbol ``*'' indicates that the constraints used are from Table \ref{AGNsummary}.}
\end{table}

\end{document}